\documentclass[preprint]{aastex}
\include{epsf}
\usepackage{longtable}

\def\gsim{~\rlap{$>$}{\lower 1.0ex\hbox{$\sim$}}}
\def\lsim{~\rlap{$<$}{\lower 1.0ex\hbox{$\sim$}}}

\def\Lsun{\hbox{$\rm\thinspace L_{\odot}$}}
\def\Msun{\hbox{$\rm\thinspace M_{\odot}$}}

\def\AJ{AJ}

\def\prd{Phys. Rev. D.}

\shorttitle{Galactic and M31 Globular Clusters : II}
\shortauthors{Beasley et al.}

\begin{document}

%% LaTeX will automatically break titles if they run longer than
%% one line. However, you may use \\ to force a line break if
%% you desire.

\title{The Chemical Properties of Milky Way and M31 Globular Clusters: II. Stellar Population Model Predictions}

%% Use \author, \affil, and the \and command to format
%% author and affiliation information.
%% Note that \email has replaced the old \authoremail command
%% from AASTeX v4.0. You can use \email to mark an email address
%% anywhere in the paper, not just in the front matter.
%% As in the title, you can use \\ to force line breaks.

\author{Michael A. Beasley}
\affil{Lick Observatory, University of California, Santa Cruz, CA 95064, USA}
\email{mbeasley@ucolick.org}

\author{Jean P. Brodie}
\affil{Lick Observatory, University of California, Santa Cruz, CA 95064, USA}
\email{brodie@ucolick.org}

\author{Jay Strader}
\affil{Lick Observatory, University of California, Santa Cruz, CA 95064, USA}
\email{strader@ucolick.org}

\author{Duncan A. Forbes}
\affil{Centre for Astrophysics \& Supercomputing, Swinburne University,
  Hawthorn, VIC 3122, Australia}
\email{dforbes@astro.swin.edu.au}

\author{Robert N. Proctor}
\affil{Centre for Astrophysics \& Supercomputing, Swinburne University,
  Hawthorn, VIC 3122, Australia}
\email{rproctor@astro.swin.edu.au}

\author{Pauline Barmby}
\affil{Harvard-Smithsonian Center for Astrophysics, 60 Garden Street, 
Cambridge, MA 02138, USA}
\email{pbarmby@cfa.harvard.edu}

%\and

\author{John P. Huchra}
\affil{Harvard-Smithsonian Center for Astrophysics, 60 Garden Street,
Cambridge, MA 02138, USA}
\email{huchra@cfa.harvard.edu}

%% Notice that each of these authors has alternate affiliations, which
%% are identified by the \altaffilmark after each name.  Specify alternate
%% affiliation information with \altaffiltext, with one command per each
%% affiliation.

\begin{abstract}
We derive ages, metallicities and abundance 
ratios ([$\alpha$/Fe]) from the integrated spectra 
of 23 globular clusters in M31, by employing multivariate fits to two 
different stellar population models.
We also perform a parallel analysis on 21 Galactic
globular clusters as a consistency check, and in order 
to facilitate a differential analysis. 
%Our derived ages and metallicities are consistent
%between stellar population models, whilst [$\alpha$/Fe] ratios
%are rather uncertain and model-dependent.
Our analysis shows that the M31 globular clusters separate
into three distinct components in age and metallicity; we identify
an old, metal-poor group (7 clusters), an old, metal-rich 
group (10 clusters) and an intermediate age (3--6 Gyr), 
intermediate-metallicity ([Z/H]$\sim$--1) group (6 clusters).
This third group is not identified in the Galactic
globular cluster sample.
The majority of globular clusters in both samples
appear to be enhanced in $\alpha$-elements, but the
degree of enhancement is rather model-dependent.
The intermediate age GCs appear to be the most
enhanced, with [$\alpha$/Fe]$\sim$0.4. These clusters are clearly 
depressed in CN with respect to the models and the bulk of the M31 and 
Milky Way sample. Compared to the bulge of M31, M32 and NGC~205, 
these clusters most resemble the stellar populations
in NGC~205 in terms of age, metallicity and CN abundance. 
We infer horizontal branch morphologies for the M31 clusters
using the Rose (1984) \ion{Ca}{2} index, and 
demonstrate that blue horizontal branches are not leading to
erroneous age estimates in our analysis.
We discuss and reject as unlikely the hypothesis that 
these objects are in fact foreground stars contaminating 
the optical catalogs.
The intermediate age clusters have generally higher
velocities than the bulk of the M31 cluster population.
Spatially, three of these clusters are projected onto the bulge region, 
the remaining three are distributed at large radii.
We discuss these objects within the context of the 
build-up of the M31 halo, and suggest that these
clusters possibly originated in a gas-rich 
dwarf galaxy, which may or may not be presently
observable in M31.
\end{abstract}

%% Keywords should appear after the \end{abstract} command. The uncommented
%% example has been keyed in ApJ style. See the instructions to authors
%% for the journal to which you are submitting your paper to determine
%% what keyword punctuation is appropriate.

\keywords{globular clusters: general -- galaxies: individual: M31}

\section{Introduction}

Hubble (1932) was the first to identify and study
globular clusters in the Andromeda galaxy (M31).
From photographic plates obtained using the 100-inch
telescope at Mount Wilson, he identified 140 objects
apparently associated with M31 for which he stated 
{\it ``...from their forms, structure, colors, 
luminosities and dimensions they are provisionally 
identified as globular clusters.''}.
From the spectrum of ``object No. 62'', he derived a radial
velocity of --210$\pm$100 kms$^{-1}$, consistent 
with the radial velocity of M31, thereby reinforcing the 
notion that these were clusters associated with the galaxy, and 
not seen in projection (nor indeed, were they background galaxies).

Despite their importance, and decades of study since Hubble, 
our knowledge about the chemical properties and ages of globular 
clusters (GCs) in M31 is limited.
Fine abundance analysis, the primary method for 
determining the chemical compositions of stars, is currently 
beyond the capabilities of current instrumentation for M31 
GCs.
Color-magnitude diagrams below the turn-off, the most reliable method
for age determinations of stellar populations, are 
extremely challenging for current instrumentation, 
and at the distance of M31 suffer from the 
dual effects of crowding and the intrinsic faintness
of cluster stars (e.g., Rich et al. 1996; Stephens et al. 2001)\footnote{In an heroic effort, 
Brown et al. (2004) have recently published a CMD to the turn-off for an M31 
GC. They derive an age of $\sim$ 10 Gyr for GC 312-035.}.
Thus, the determination of fundamental parameters such 
as metallicity have relied upon studies
of their integrated light (e.g., van den Bergh 1969; 
Burstein et al 1984; Battistini et al. 1987; 
Tripicco 1989; Huchra et al. 1991; Jablonka et al. 1992;
Jablonka et al. 1998; Ponder et al. 1998; 
Barmby et al. 2000; Perrett et al. 2002; Jiang et al. 2003.)

The general picture which has emerged 
from studies such as these is that M31's GCs are broadly
similar to their counterparts in the Milky Way.
Both the Milky Way and M31 GC system metallicity ([Fe/H])\footnote{Unless
otherwise stated, [Fe/H] refers to metallicities on
the Zinn \& West (1984) scale. [Z/H] refers
to 'total' metallicity, and [$\alpha$/Fe] represents
the logarithmic ratio of $\alpha$-capture elements 
(e.g., O, Mg, Ca...) to iron.} distributions
are consistent with being bimodal, with peaks
at [Fe/H]$\sim$--1.5 and $\sim$--0.5 
(Huchra et al. 1991; Barmby et al 2000; Perrett 2004). 
The most notable difference
in terms of the chemical properties of the 
two cluster systems is that CN in the M31 clusters 
appears to be significantly enhanced with respect 
to Milky Way GCs (Burstein et al. 1984; Brodie \& Huchra 1991; 
Tripicco 1989; Beasley et al. 2004; henceforth Paper I). 
On the basis of stellar modelling 
(Tripicco \& Bell 1992) and observations of the near-UV NH feature
(Ponder et al 1998; Li \& Burstein 2003), the interpretation is that N is significantly
enhanced in the M31 GCs.

The majority of GCs in M31 appear to be
old, although there have been  
claims for the existence of a few intermediate age clusters 
(Jablonka et al. 1998; Jiang et al. 2003; Trager 2004). 
Based upon optical and near-IR colors, 
Barmby \& Huchra (2000) have claimed that 
the ensemble of metal-rich GCs in M31 may be up to
4 Gyr younger than their metal-poor counterparts.
In Paper I we showed that
a number of star clusters in M31, identified as having
thin-disk like kinematics (Morrison et al. 2004) are 
young objects ($<$1 Gyr old) with near-solar metallicities.
We refer to these objects as 'star clusters' rather
than GCs since, in the absence 
of spectroscopic mass estimates for these clusters, it is 
presently unclear whether they possess true globular
cluster masses (e.g., Larsen et al. 2004), or are more akin
to open clusters\footnote{This terminology for star clusters
may be a semantic point. However, it is important to 
recognize that no star clusters of globular cluster mass 
$\gsim 10^4\Msun$ are thought to be associated with the Galactic thin disk.}.

Notwithstanding, the majority of these studies, in particular 
the spectroscopic ones, have by necessity been rather 
qualitative in nature, since only
recently have stellar population models been available 
which can account for a range of metallicities, ages and $\alpha$-element
abundance ratios (see Trager et al. 2000; Proctor \& Sansom 2002; 
Thomas, Maraston \& Bender 2003; Proctor, Forbes \& Beasley 2004; hereafter PFB04).
In this paper, we compare the Galactic and M31 GC samples
presented in Paper I with contemporary stellar population
models to estimate their ages, metallicities, and [$\alpha$/Fe]
ratios.

This paper is organized in the following fashion.
In Section~\ref{Data} we briefly summarize the data
used in this study. In Section~\ref{NSARS} we
discuss the models and their correction
for non-solar abundance ratios.
In Section~\ref{Analysis}, we discuss
our methodology and derive fundamental 
parameters (age, metallicity, [$\alpha$/Fe])
for the Milky Way and M31 GCs. 
The physical nature of six intermediate
age cluster candidates in M31 are discussed
in Section~\ref{IAGC}.
Section~\ref{HBR} presents an estimation
of the horizontal branch morphologies 
of the M31 GCs based upon integrated
spectra. 
Finally, in Section~\ref{Summary}, 
we discuss the results of this study.

\section{Galactic and M31 Data}
\label{Data}

The Milky Way and M31 GC data used here were discussed
in detail in Paper I. To summarize, we use 20 high-quality 
integrated spectra for Milky Way globular clusters
taken from Puzia et al. (2002; hereafter P02) and Cohen, 
Blakeslee \& Ryzhov (1998; hereafter CBR98).
The M31 data were first discussed in Barmby et al. (2000), 
and were re-analysed following procdures given in Paper I.
Here we analyze 23 M31 GCs, excluding the young disk
objects identified in Paper I.
Unless otherwise stated, the P02 and CBR98+M31 data have been 
corrected onto the Lick spectroscopic system using the offsets
given in P02 and Paper I respectively. All the Lick indices used
here employ the Trager et al. (1998) definitions (supplemented
by the higher order Balmer lines given by Worthey \& Ottaviani 1997), 
which provide greater consistency with the Worthey (1994) 
fitting functions than the definitions given in Worthey et al. (1994)
(G. Worthey, private comm.).
These data have been augmented by an integrated spectrum
of the outer bulge of the Milky Way (P02) and 
integrated spectra for the central bulge of M31 (r$_e$/8 aperture), 
and two of its companions M32 (NGC~221) and NGC~205 (Trager et al. 1998).

\section{Stellar Population Models for Non-solar Abundance Ratios}
\label{NSARS}

It has been known for some time that elliptical galaxies (and 
Milky Way GCs) exhibit [$\alpha$/Fe] ratios which differ from
the local abundance pattern (e.g. Worthey, Faber \& Gonz$\acute{a}$lez 1992).
In order to quantify the metallicities and ages 
of such systems, various attempts have been made to either empirically
or theoretically calibrate models which account for non-solar
abundance patterns (e.g. Barbuy 1994; Borges et al. 1995; Milone et al. 2000). 
However, only recently 
have comprehensive models been available
which can account for a wide range of metallicity, age and 
$\alpha$-element abundances in the commonly used
Lick system (Trager et al. 2000; 
Proctor \& Sansom 2002; Thomas, Maraston \& Bender 2003; 
Tantalo et al. 2004).

In this paper, we considered three different single stellar population
(SSP) models, those of  Vazdekis (1999; henceforth V99), 
Bruzual \& Charlot (2003; henceforth BC03) and 
those of Thomas, Maraston \& Bender (2003; henceforth TMB03). 
The V99, BC03 and TMB03 models provide indices based on Lick fitting
functions, which define relations between index strength and metallicity, 
effective temperature and gravity. 
In addition, V99 and BC03  make available spectral energy distributions
which allow flexibility in defining and measuring 
line-strength indices. Here we have used
indices based upon the fitting-functions to retain 
greater consistency among the models. 

The V99 and BC03 models do not take into account
non-solar abundance ratios, and we have attempted to do so
using the method described in Trager et al. (2000).
Note that we use the Trager et al. (2000) method, which 
assumes a fixed [Z/H], rather than that discussed
by Proctor \& Sansom (2002), which assumes a fixed [Fe/H], 
since the former is more appropriate for metal-poor
stellar populations (Proctor \& Sansom 2002).
Briefly, the corrections referred to above require a knowledge
of the following:
{\it (i)} the abundance ratio
pattern of the stars in the SSP model stellar library, {\it (ii)}
the difference between this pattern and 
objects in question (in this case GCs), {\it (iii)} 
the impact of this difference on the model isochrones
used to construct the SSP and {\it (iv)} the change in 
line-strength indices due to a given change in the abundance
ratio pattern for a given isochrone (synthetic 'response'
functions). There are a number of caveats 
and limitations to these corrections; e.g., point {\it (i)} has yet
to be homogenously characterised for any stellar library, whilst 
the Tripicco \& Bell (1995) response functions required in 
point {\it (iv)} were only performed for a very limited parameter
space (three stars), with what are now rather old atomic
linelists.
For a  full discussion of the treatment of varying abundance ratios in the
models, see Trager et al. (2000), Proctor \& Sansom (2002), Proctor et al. (2004) 
and PFB04.

The TMB03 models already account for 
variations in [$\alpha$/Fe], and now include corrections for the
higher-order Balmer lines (Thomas, Maraston \& Korn 2004; henceforth TMK04)
which were not modeled by Tripicco \& Bell (1995).
For all three models, we find reasonable consistency between 
their age, metallicity and [$\alpha$/Fe] predictions for 
the GC data (see PFB04 and Section~\ref{Multivariate}).
For this reason, and due to the fact that the V99 models do not extend to 
metallicities below [Fe/H]=--1.7, we only utilise the BC03 and 
TMB03+TMK04 models (these latter two models 
referred to collectively as TMB03) in what follows.

\section{Analysis}
\label{Analysis}

In order to characterize where the M31 and Milky Way GCs 
lie in the parameter space of the models, we begin by examining 
the behaviour of some individual indices. 
In Section~\ref{Multivariate}, we perform more quantitative 
multivariate fits to identify which modeled combinations
of age, metallicity and [$\alpha$/Fe] best describe these
data. To previse some of the results of our age-metallicity analysis, 
six  M31 GCs (including 292-0101, an intermediate age cluster
identified in Paper I) appear to be of intermediate age 
($\sim$ 5 Gyr) and have
intermediate metallicities ([Z/H]$\sim$--1.0). These
GCs are indicated as such in the following Section, 
and the justification for this is given in  Section~\ref{Multivariate}.

\subsection{Index-Index Comparisons}
\label{Index}

In Figure~\ref{TMBMgbFe} we plot $\langle$Fe$\rangle$\footnote{
(Fe5270+Fe5335)/2}
versus Mg $b$ for the M31 and Milky Way GCs, compared to the 
TMB03 models. This combination of indices has been shown to 
possess differing sensitivities to Fe-peak and $\alpha$-element
abundance ratios (Tripicco \& Bell 1995), and can therefore
be used to track the variation of [$\alpha$/Fe]
(Trager et al. 2000; TMB03). 
As already demonstrated by TMB03, and can be seen 
in Figure~\ref{TMBMgbFe}, the Milky Way data  
occupy a position at approximately
[$\alpha$/Fe]=+0.3, which is roughly consistent with high-dispersion
estimates for Milky Way GCs (e.g., see the compilation of Carney 1996; 
Gratton et al. 2004).
Based on this figure, the M31 GCs also appear to be enhanced in 
$\alpha$-elements,  
but at the level of [$\alpha$/Fe]=+0.2 for Mg $b>$ 1.5, slightly less 
than that seen in the Milky Way clusters (see also Kuntschner et al. 2002
and Trager 2004). At Mg $b<$1.5, the TMB03 models suggest 
[$\alpha$/Fe]$\leq$0, behavior which is also hinted at 
for the CBR98 data but not for that of P03.
Whether this is a failure of the models, or a calibration
problem in the M31 and CBR98 data is unclear at present.

Assuming that the M31 clusters have comparable ages
to their Galactic counterparts (an as yet untested assumption), 
then based on Figure~\ref{TMBMgbFe} we may conclude
that the M31 GCs are $\alpha$-enhanced in a broadly similar fashion to 
that seen in Milky Way GCs and massive elliptical galaxies (e.g. Carney 1996; 
Trager et al. 2000)\footnote{We note that de Freitas Pacheco (1997) concluded that the
12 M31 GCs from Burstein et al. (1984) exhibited [Mg/Fe] =0.35$\pm$0.10, 
which is consistent with our findings. Unfortunately, this study does not
go into sufficient detail to allow a comprehensive appraisal.}.
The bulges of both the Milky Way and M31 also appear to be $\alpha$-enhanced 
at the 0.2 dex level, based on Figure~\ref{TMBMgbFe}, consistent 
with previous findings (e.g., McWilliam \& Rich 1994 and Trager et al. 2000
respectively). 
The position of the compact elliptical M32, suggests that this galaxy
has [$\alpha$/Fe] ratios closer to the solar value, whereas that of
the dE NGC~205 suggests [$\alpha$/Fe]$<$0.0.

\vbox{
\begin{center}
\leavevmode
\hbox{%
\epsfxsize=10cm
\epsffile{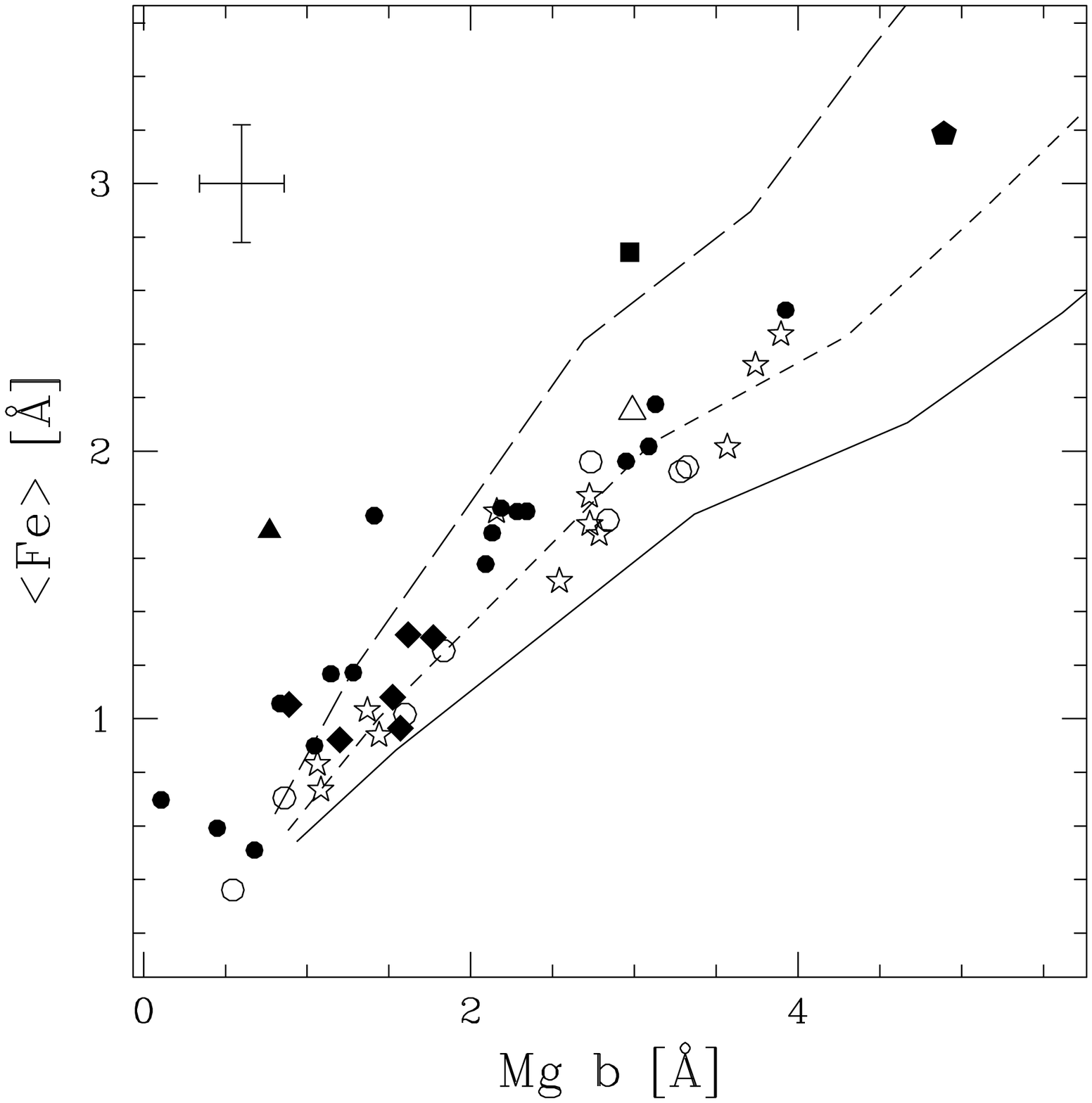}}
\figcaption{\small
Predictions of the Thomas, Maraston \& Bender (2003) 
stellar population models 
for various [$\alpha$/Fe] ratios versus the Milky Way and 
M31 GC data. Solid circles represent M31 GCs, open
symbols Milky Way GCs (stars:P02; circles:CBR98).
Models are for 12 Gyr stellar populations, 
with [$\alpha$/Fe]=0.5, 0.3 and 0.0 (solid, short-dashed and
long-dashed lines respectively).
Also shown are the Milky Way bulge data from P02 (open triangle), 
and data for the M31 bulge, M32 and NGC~205 from
Trager et al. (1998) (filled pentagon, square and triangle
respectively). The median error of the M31 sample is 
shown in the top-left of the figure.
The large diamonds are M31 clusters which we identify
as intermediate age.
\label{TMBMgbFe}}
\end{center}}

For several combinations of indices such as those 
displayed in Figure~\ref{TMBMgbFe}, consistent 
results are obtained for the GC [$\alpha$/Fe] ratios.
This is true for some  
combinations of Lick 'iron' indices (e.g., Fe4383, 
Fe5270, Fe5335, Fe5709), and 'magnesium'
indices (Mg$_2$ \& Mg $b$).
However, for a number of indices the models do not
well describe the behaviour of the Galactic GC data
(TMB03), and this is also true of the M31 GCs.
It is not immediately clear why the index
combination in Figure~\ref{TMBMgbFe} should
be preferred over other permutations of indices, 
other than its ability to fit the 
Galactic GC data. This is one reason why we prefer
the multivariate approach described in Section~\ref{Multivariate}.

\vbox{
\begin{center}
\leavevmode
\hbox{%
\epsfxsize=10cm
\epsffile{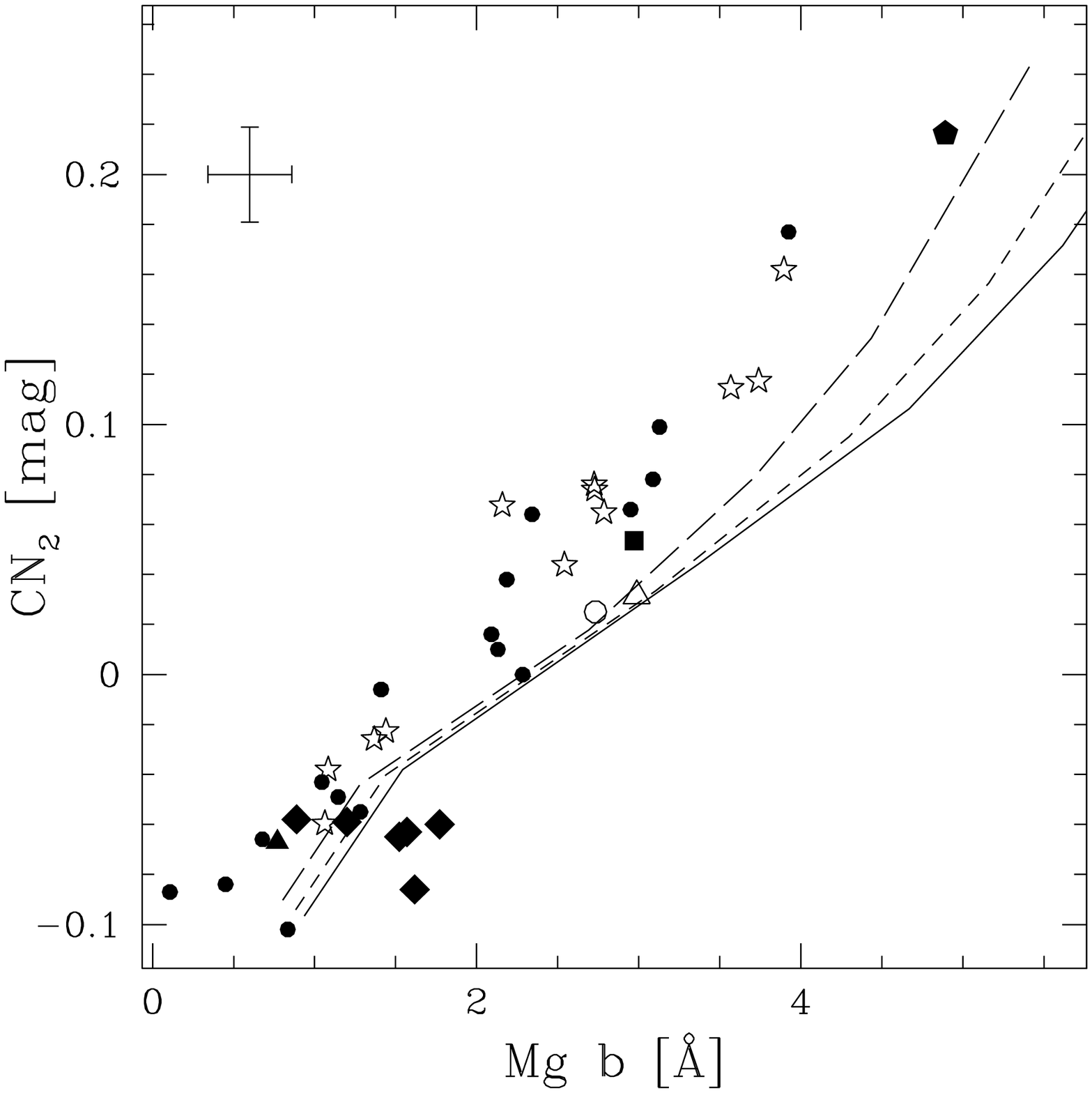}}
\figcaption{\small
Mg $b$ and CN$_2$ indices for the M31 and Milky Way GCs, 
plotted with the predictions of the Thomas, Maraston \& Bender (2003) 
stellar population models.  
Models are for 12 Gyr stellar populations, 
with [$\alpha$/Fe]=0.5, 0.3 and 0.0 (solid, short-dashed and
long-dashed lines respectively).
Symbols are as for the previous figure, but note that the CBR98 data are
not included since they do not extend to CN$_2$.
The CN$_2$ indices of the GCs and M31 bulge are clearly too strong for the 
models. The M31 GCs identified as intermediate age (diamonds) appear
{\it depressed} in CN$_2$ with respect to the models and 
bulk of M31 and Milky Way clusters.
\label{TMBCN2}}
\end{center}}

Three of the discrepant indices include the Lick CN indices
and Ca4227, which are affected by variations
in N and C (e.g., Tripicco \& Bell 1995; 
Vazdekis et al. 1997; Schiavon et al. 2002).
By way of illustration, we show the Lick CN$_2$ 
band versus Mg $b$ in Figure~\ref{TMBCN2}.
The CN$_2$ (and the CN$_1$ band - not shown)
is generally too strong compared to the models for both the Milky Way
clusters (see TMB03) and M31. The existence of strong CN enhancement
in M31 GCs when compared to the local abundance pattern, 
interpreted as N enhancement, has been known 
for some time (Burstein et al. 1984; 
Brodie \& Huchra 1991; Ponder et al. 1998; 
Li \& Burstein 2003; Burstein et al. 2004).
The M31 GCs in Figure~\ref{TMBCN2} do not appear enhanced in CN compared 
to the Milky Way clusters, although note that the Lick
correction for the M31 data is rather uncertain 
and large ($\sim$0.03 mag, see Paper I).
However, the near-UV cyanogen feature at $\lambda$3883 does appear enhanced
in M31 clusters when compared to their Galactic counterparts (Paper I), 
possibly suggesting flux-calibration problems with the Lick CN index.
The bulge of M31 also appears to be quite CN-strong, which is not
the case for (the more metal-poor) Galactic bulge.
TMB03 found that they could fit the CN indices of the Milky Way 
data by assuming [$\alpha$/N]=--0.5, or in other words by increasing
N by a factor of 3 with respect to the $\alpha$-elements.
Presumably, the M31 GCs are enhanced in N by at least a similar factor. 

\vbox{
\begin{center}
\leavevmode
\hbox{%
\epsfxsize=10cm
\epsffile{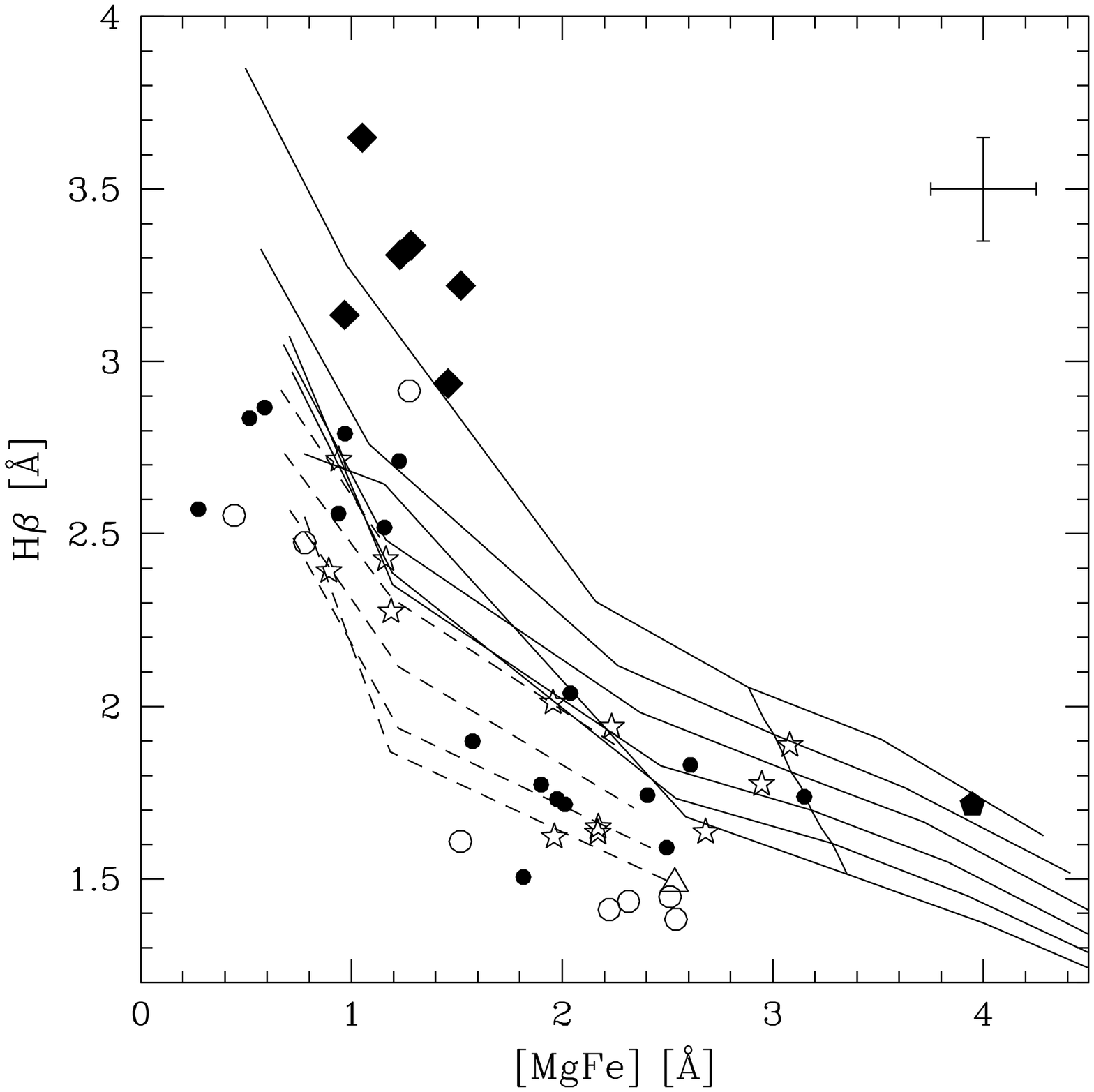}}
\figcaption{\small
The H$\beta$ indices as a function
of [MgFe] for the M31(filled circles) and Milky Way GCs (open circles: CBR98; 
open stars: P02). Overplotted are the Thomas, Maraston \& Korn (2004) models
assuming [$\alpha$/Fe]=+0.3 for ages 5,7,9,11,13 \& 15 Gyr (solid lines).
The single angled line at [MgFe]$\sim$3 represent an isometallicity
line of [Z/H]=0.0. The dashed lines represent the models of 
Maraston \& Thomas (2000) for ages 9,11,13 \& 15 Gyr at [Z/H]$\leq$--0.5, 
which assume no mass-loss on the red giant branch (see text).
Six M31 clusters (diamonds) have clearly enhanced H$\beta$. 
\label{hbeta}}
\end{center}}

\vbox{
\begin{center}
\leavevmode
\hbox{%
\epsfxsize=10cm
\epsffile{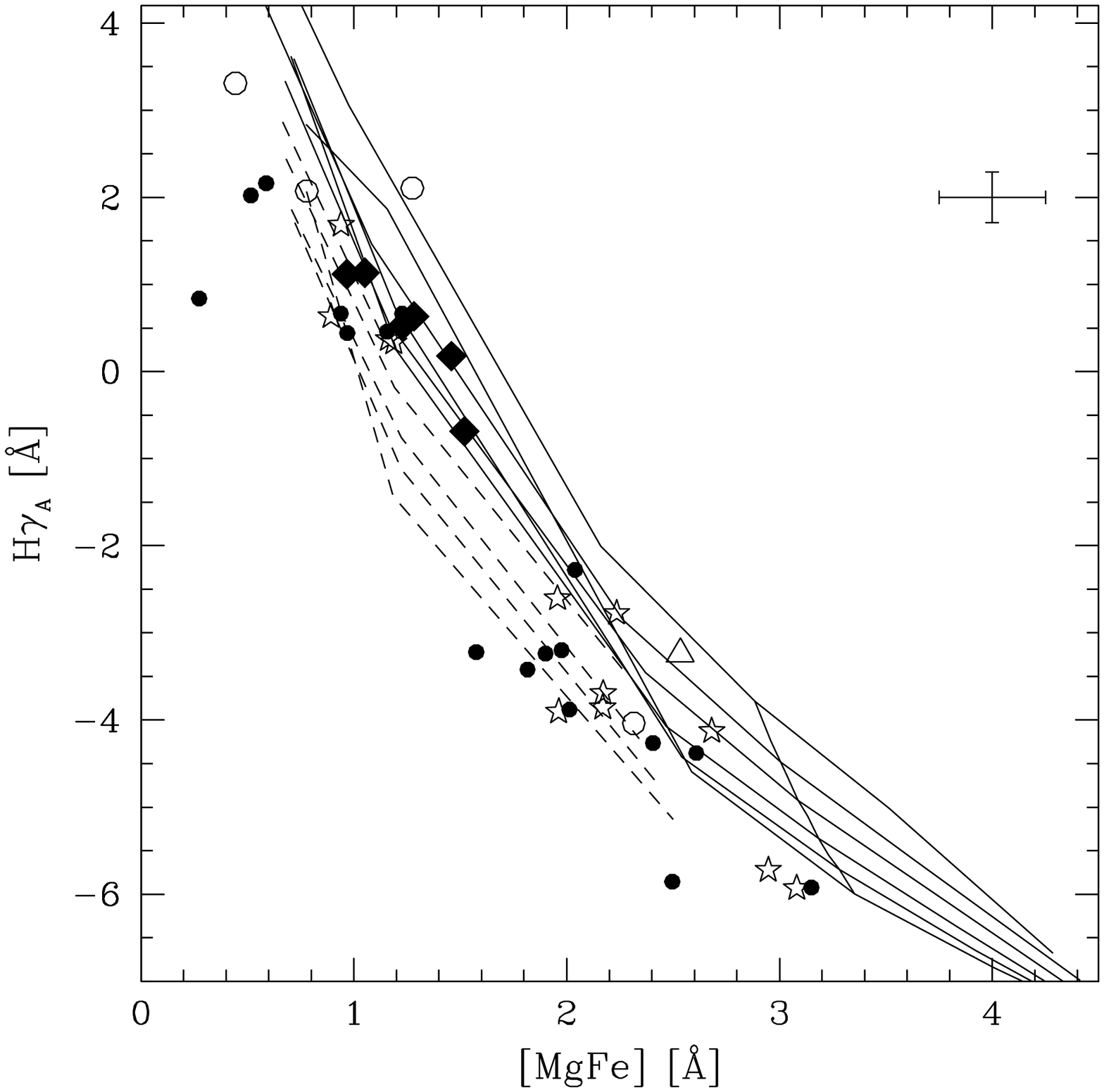}}
\figcaption{\small
The H$\gamma_{\rm A}$ indices as a function
of [MgFe]. Symbols as for previous figure. 
\label{hgamma_a}}
\end{center}}

\vbox{
\begin{center}
\leavevmode
\hbox{%
\epsfxsize=10cm
\epsffile{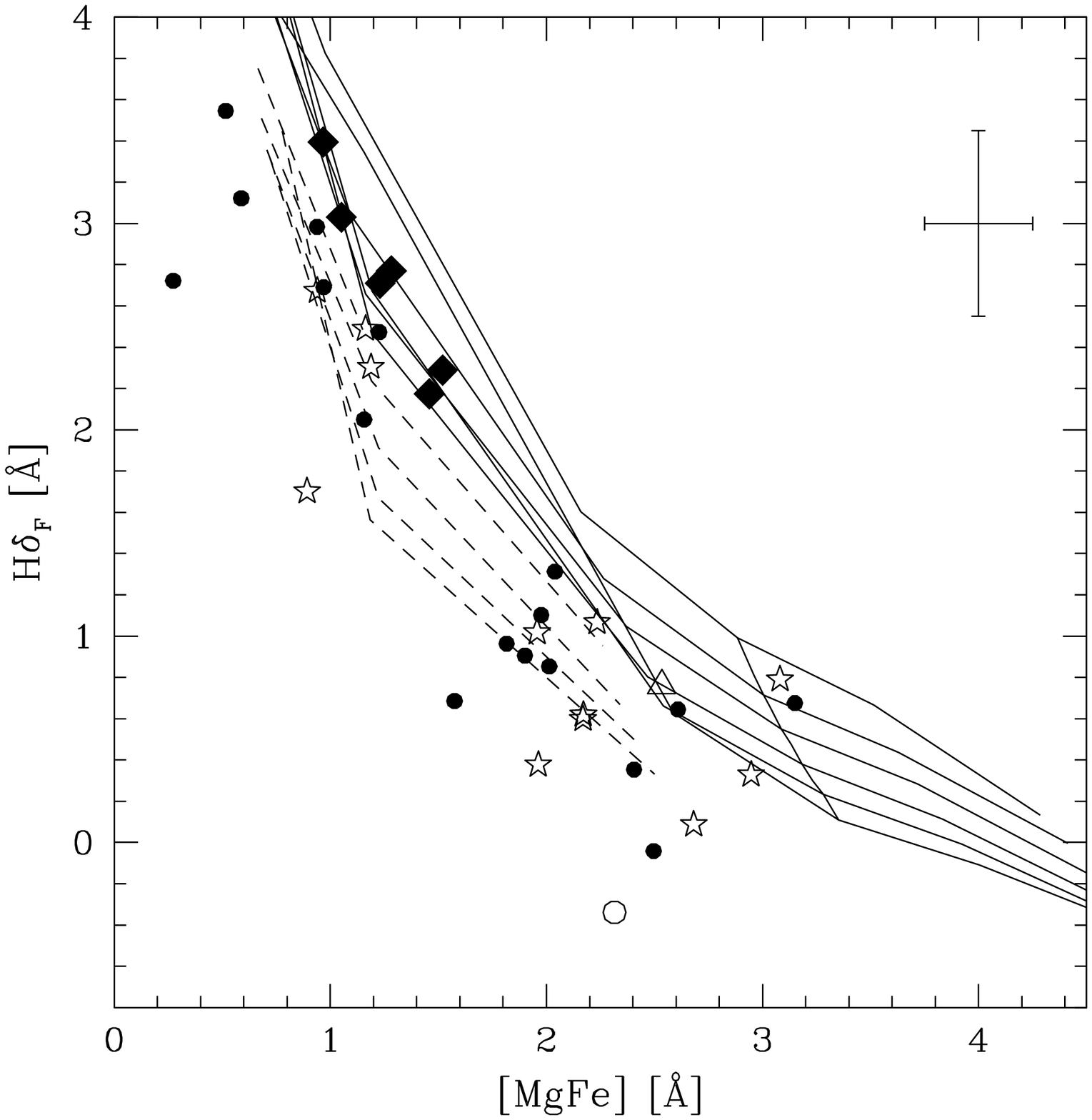}}
\figcaption{\small
The H$\delta_{\rm F}$ indices as a function
of [MgFe]. Symbols as for Figure~\ref{hbeta}.
\label{hdelta_f}}
\end{center}}

The positions of the Milky Way and M31 GCs in the H$\beta$--[MgFe]
planes of the TMK04 models are shown in Figure~\ref{hbeta}.
Here, we have adopted the [$\alpha$/Fe]=+0.3 models of TMK04, 
consistent with the location of these data in Figure~\ref{TMBMgbFe}.
These data span a wide range of metallicities, approximately
--2.0$\leq$[Z/H]$\leq$0.
Figures~\ref{hgamma_a} and \ref{hdelta_f} show these data 
in the H$\gamma_{\rm A}$--[MgFe] and H$\delta_{\rm F}$--[MgFe]
planes respectively.
Both the Milky Way and M31 GCs generally define similar, old-age sequences
in H$\beta$, H$\gamma_{\rm A}$ and H$\delta_{\rm F}$ (we have not used
H$\gamma_{\rm A}$ since this index showed large calibration
uncertainties in the M31 data; see Paper I). Many of the data points 
fall off the bottom of the TMK04 grids, suggesting that only lower limits may 
be derived for many clusters from these diagrams.
The dashed lines indicate the models of Maraston \& Thomas (2000), 
which assume generally red horizontal branches, and have not been corrected for the 
local [$\alpha$/Fe] abundance pattern\footnote{At low metallicities, 
these models actually reflect an $\alpha$-enhanced pattern following
Galactic halo stars (TMB03).}.
These model variants do a much better job of following the data, 
suggesting rather modest mass-loss on the red giant branch
(hence, redder horizontal branches) for stars in these clusters.

We identify five M31 clusters (126-184, 301-022, 337-068, NB16, and NB67)  
as intermediate age and of intermediate metallicity.
%an assertion which is justified in Section~\ref{Multivariate}.
They appear particularly anomalous in the H$\beta$--[MgFe] plot, 
but appear more 'normal' in their higher-order Balmer lines.
Closer inspection of Figures~\ref{hgamma_a} and \ref{hdelta_f}  suggest
that these objects also define a different locus 
in H$\gamma_{\rm A}$ and H$\delta_{\rm F}$. These objects lie in a region
very similar to the metal-poor, old GCs and without the 
benefit of SSP models were not identified as being particularly unusual
in Paper I.

\vbox{
\begin{center}
\leavevmode
\hbox{%
\epsfxsize=10cm
\epsffile{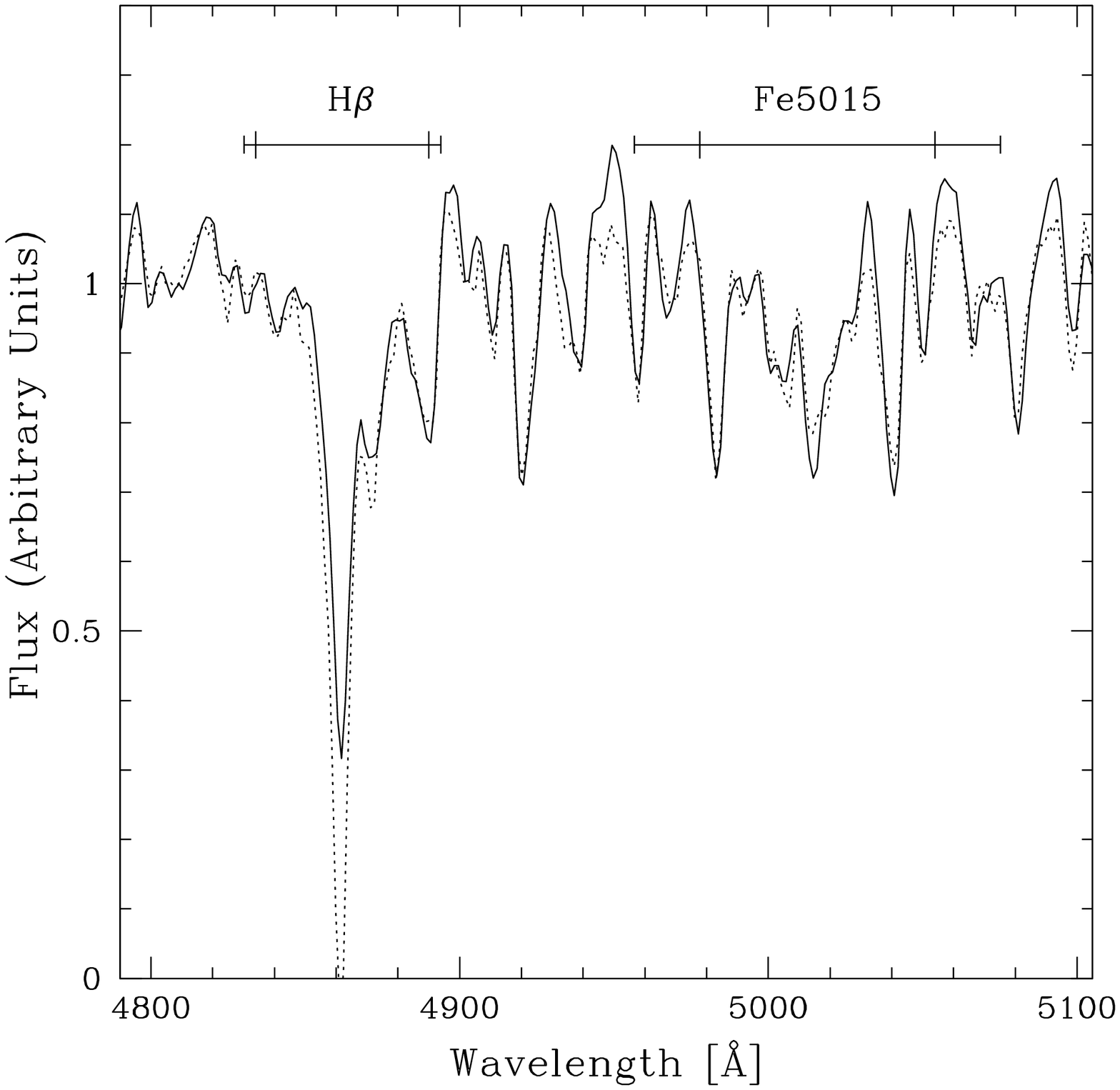}}
\figcaption{\small
A comparison of the spectra
of two GCs, 158-213 and 337-068 in M31, which 
have similar metallicities but different ages.
The spectra are shown in a narrow wavelength
range to illustrate important similarities and differences
between the two spectra. The index and continuum passbands of the Lick
H$\beta$ and Fe5015 indices are marked. 
The solid line indicates the spectrum of 158-213, an
old GC (12.1$\pm$0.9 Gyr) with [Z/H]=--0.8$\pm$0.2
(Thomas, Maraston \& Bender 2003 models).
The dotted line shows 337-068, an IAGC (6.6$\pm$3.2 Gyr) 
with [Z/H]=--1.0$\pm$0.2.
Note the fairly good agreement between the metallicity-sensitive
Fe5015 index, but the significantly stronger H$\beta$
absorption in the younger GC.
\label{spectra}}
\end{center}}

To illustrate the differences in the H$\beta$ indices
between the intermediate age and old GCs, a comparison
of the spectra of the M31 GCs 158-213 and 337-068
is shown in Figure~\ref{spectra}.
These clusters are at roughly similar metallicities
([Z/H]$\sim$--1.0), but cluster 337-068, which 
we identify as being intermediate age, 
has significantly stronger H$\beta$ absorption.
The similar metallicities of these clusters
is confirmed by comparison of the Fe5015 index
(the slightly weaker Fe5015 index in 337-068
with respect to 158-213 can be explained purely
by the age, rather than metallicity difference
between these clusters).
Strong Balmer-lines in M31 GCs have been noted previously
(Spinrad \& Schweizer 1972; Rabin 1981; Burstein et al. 1984)
and are not in themselves sufficient evidence for 
young ages. Stellar 
populations such as hot horizontal branch stars
can also contribute to absorption in the Balmer
lines. In the next section we justify our
identification  of intermediate ages for these
clusters.

\subsection{Model Comparisons through Multivariate Fitting}
\label{Multivariate}

Based upon diagnostic plots of individual indices 
such as shown in Figure~\ref{hbeta}, age, metallicity and abundance 
ratio determinations of the GCs in our sample are often ambiguous. 
The combination
observational errors (systematic+random), and modeling uncertainties
in the Balmer lines (e.g., due to horizontal branch morphology variations)
make age estimations particularly uncertain. This is well
illustrated in Figures~\ref{hbeta}, \ref{hgamma_a} and
\ref{hdelta_f} for the  
most metal-rich Galactic GC in this study, NGC~6553 (from P02).
The H$\beta$ index predicts an age of $\sim$ 7 Gyr for this cluster, 
H$\gamma_{\rm A}$ predicts ages $\geq$15 Gyr, while H$\delta_{\rm F}$
yields $\sim$ 6 Gyr. Although in the absence of a preference of one 
index over another, a weighted mean age may be appropriate, 
this is an appreciable age-range for what is thought
to be a $\sim$ 13 Gyr cluster from its CMD (Zoccali et al. 2001).

Therefore, we have opted to use the method outlined in PFB04, 
and utilized in Paper I\footnote{We have elected not to 
list derived [$\alpha$/Fe] ratios for the young disk clusters
identified in Paper I, since the model predictions are 
highly uncertain at such young ages. However, we note that
the Mg $b$/$\langle$Fe$\rangle$ ratios of these
objects are consistent with solar-scaled
[$\alpha$/Fe] ratios.} and Pierce et al. (2004, MN submitted).
PFB04 implement a straightforward $\chi^2$ minimization
technique between measured Lick indices and SSP models, in order to identify the 
best model solution for age, metallicity and [$\alpha$/Fe].
This method, in somewhat different forms, has been successfully used for galaxies
(e.g., Vazdekis et al. 1997; Proctor \& Sansom 2002) and GCs
(e.g., de Freitas Pacheco 1997, -- see also Barmby \& Huchra 2000).
In essence, the entire suite of measured Lick indices
are compared to the full grid of model indices, which 
are searched for the best solution for the above parameters. 
Any indices which significantly deviate (by $>$ 3 $\sigma$) are
clipped from the fit and $\chi^2$ re-calculated. This procedure is
iterated until no further indices are removed and a stable solution 
obtained. As demonstrated by PFB04, the method is robust, and 
recovers reliable ages, metallicities and abundance ratios for 
Galactic GCs using the V99, BC03 and TMB03 models.

\vbox{
\begin{center}
\leavevmode
\hbox{%
\epsfxsize=10cm
\epsffile{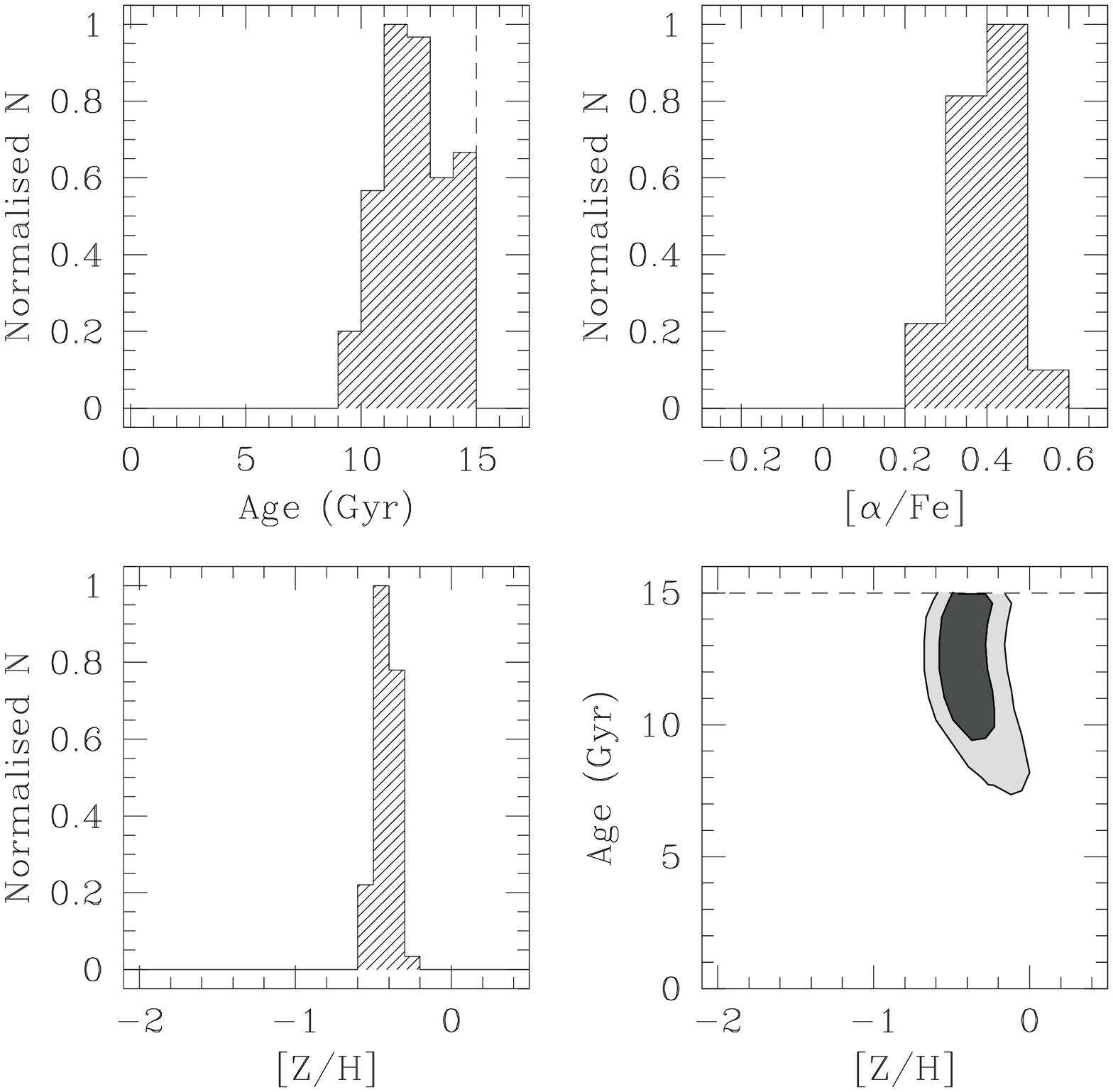}}
\figcaption{\small
The 1-$\sigma$ confidence distributions for age, metallicity and 
[$\alpha$/Fe] for NGC~6536, after performing a $\chi^2$ fit 
to the Thomas, Maraston \& Bender (2003) SSP models. 
The bottom-right panel shows the 1-$\sigma$
and 2-$\sigma$ confidence regions in the [Z/H]--Age plane.
The banana-shape of this region illustrates that age
and metallicity are still degenerate in the solution.
The dashed lines in the top-left and bottom-right
panels indicate the upper age-limit (15 Gyr) in the models.
The piling up in the oldest bin in the top-left panel is
artificial.
\label{chi}}
\end{center}}

An example of the application of this method to the moderately metal-rich
Galactic GC NGC~6536 (P02 data) is shown in Figure~\ref{chi}.
The best-fit parameters for this GC, based upon the TMB03 (BC03) models, 
are [Z/H]=--0.4$\pm$0.1 (--0.7$\pm$0.1), [$\alpha$/Fe]=0.4$\pm$0.1 (0.2$\pm$0.2) 
and an age of 13$\pm$2 (11$\pm$3) Gyr.
Since the three-dimensional error distributions in the derived 
parameters are generally asymmetric, uncertainties
are taken to correspond to the maximum difference of the parameter
in question on the 1-$\sigma$ surface. 
Using this fitting procedure, we have derived
age, [Z/H] and [$\alpha$/Fe] for the Galactic GC data (P02 and CBR98), 
and the M31 clusters. The best solutions for age, [Z/H] and [$\alpha$/Fe] 
for the M31 GCs, using the BC03 and TMB03 models, are given
in Table~\ref{Solutions}.
As a result of the $\chi^2$ fits, several indices were routinely
clipped from these datasets, indicating aberrant index values
{\emph with respect to the stellar population models in question}. 
Specifically, for the M31 data the NaD index was always 
rejected, and we attribute this to problems with interstellar 
absorption in this index (e.g., Burstein et al., 1984; TMB03). 
Furthermore, generally both the CN$_1$ and CN$_2$ indices were
significantly elevated with respect to the model predictions
(see later). In the case of the Milky Way data, again the CN and 
NaD indices were rejected, as was the Fe5015 index in the P02 data 
which appears to have calibration issues (see P02 and PFB04).
There are four common GCs between 
the CBR98 and P02 data, and we find encouraging consistency
in our solutions between these objects (PFB04). In the following, we
use the solutions for the P02 rather than CBR98 clusters in
common, since the P02 data generally have more indices 
available to fit. We refer to these ensemble data as the 
'Galactic GCs'.

\vbox{
\begin{center}
\leavevmode
\hbox{%
\epsfxsize=16cm
\epsffile{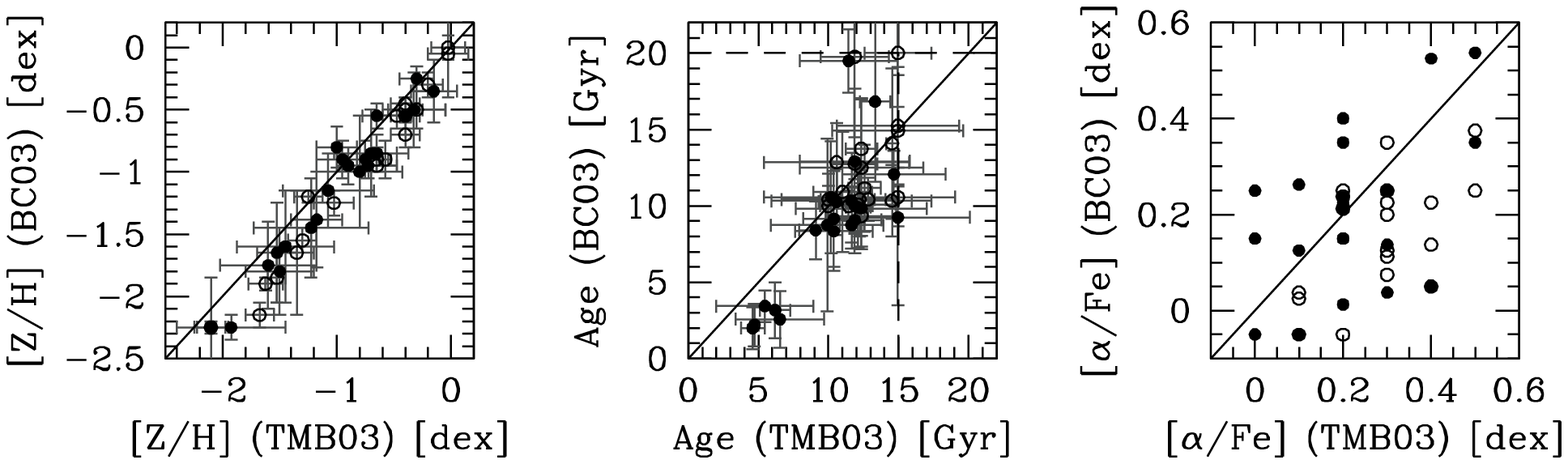}}
\figcaption{\small
Comparison between derived parameters for the Galactic (open circles) 
and M31 GCs (filled circles) using the
Bruzual \& Charlot (2003) and Thomas, Maraston \& Bender (2003)
models. The latter models incorporate higher-order Balmer lines 
from Thomas, Maraston \& Korn (2004).
In each panel, the solid line represents unit slope
for reference. Error bars have been omitted from the far-right
panel for clarity. In general, the agreement between models
is acceptable. 
\label{compare_models}}
\end{center}}

Differences between SSP models can be significant, reflecting
differing choices made by modelers on issues such as 
the stellar library, the initial mass function and the 
choice of isochrone. 
In Figure~\ref{compare_models} we compare the best solutions 
using the PFB04 method for the BC03 and TMB03 models
applied to the Galactic and M31 GCs.
The good correlation in metallicity between the BC03 and 
TMB03  models is very encouraging. However, there is clearly a 
deviation from unit slope. At the most metal-poor end, 
the TMB03 models predict metallicities $\sim$ 0.25 dex
higher than the BC03 models. This offset decreases
at higher metallicities, until no net offset 
is seen at [Z/H]$\sim$0.
In terms of ages, the agreement is also encouraging.
Five M31 clusters clearly lie at relatively young ages, 
$\sim$ 6 (3) Gyr according to the TMB03 (BC03) models.
These clusters all have similar metallicities,  
with [Z/H]$\sim$--1.0.
A number of GCs also pile-up on the oldest-age line of the 
TMB03 models, suggesting that their age solutions
lie 'older' than the models allow. This problem is not seen 
in the BC03 models, which have a larger age range, 
although absolute ages of 20 Gyr are inconsistent 
with the presently favored age of the Universe 
(Tegmark et al. 2004).

The agreement between the [$\alpha$/Fe] ratios is
less convincing, although a correlation is present.
The BC03 models predict [$\alpha$/Fe] ratios $\sim$ 0.15 dex
lower than TMB03 for the Galactic GC data.
In correcting for the local abundance pattern in the BC03
models, we have assumed the same pattern
as that adopted by TMB03, which shifts from 
[$\alpha$/Fe]=+0.25 at [Z/H]=--2.25 to 
[$\alpha$/Fe]=0 at [Z/H]=0. 
This is quite a low value to adopt, when one
considers that [$\alpha$/Fe]=0.4$\sim$0.6 is more in keeping 
with the local stellar abundance pattern of (for example) [O/Fe], [Mg/Fe]
and [Ca/Fe] at [Fe/H]$\leq$--1 
(see Gratton et al. 2004 and references
therein). The [$\alpha$/Fe] pattern adopted by TMB03
reflects the mean of [Mg/Fe] and [C/Fe]
observed in the solar neighborhood, which 
are $\sim$ 0.4 and $\sim$ 0.1 respectively (at [Fe/H$<$--1), and
therefore presumably reflects [Mg+C/Fe].
To maintain consistency with TMB03, we
adopt their values for this correction.

In summary, there is generally acceptable agreement
between the predictions of the BC03 and TMB03 models.
Of the three key parameters we are interested in, the
reliability of the predicted [$\alpha$/Fe] ratios are the 
least convincing. Since the derived parameters are still
somewhat model-dependent, we will explicitly state which
model was used when quoting values for age, metallicity
and abundance ratios.
We emphasize that the ages, metallicities and abundance ratios
discussed in the following sections do not explicitly take into
account the intrinsic uncertainties in the stellar population
models used. Moreover, absolute values, particularly for ages, 
should be treated with caution.

\subsection{Globular Cluster Ages, Metallicities and Abundance Ratios}
\label{AgesandMetallicities}

\vbox{
\begin{center}
\leavevmode
\hbox{%
\epsfxsize=10cm
\epsffile{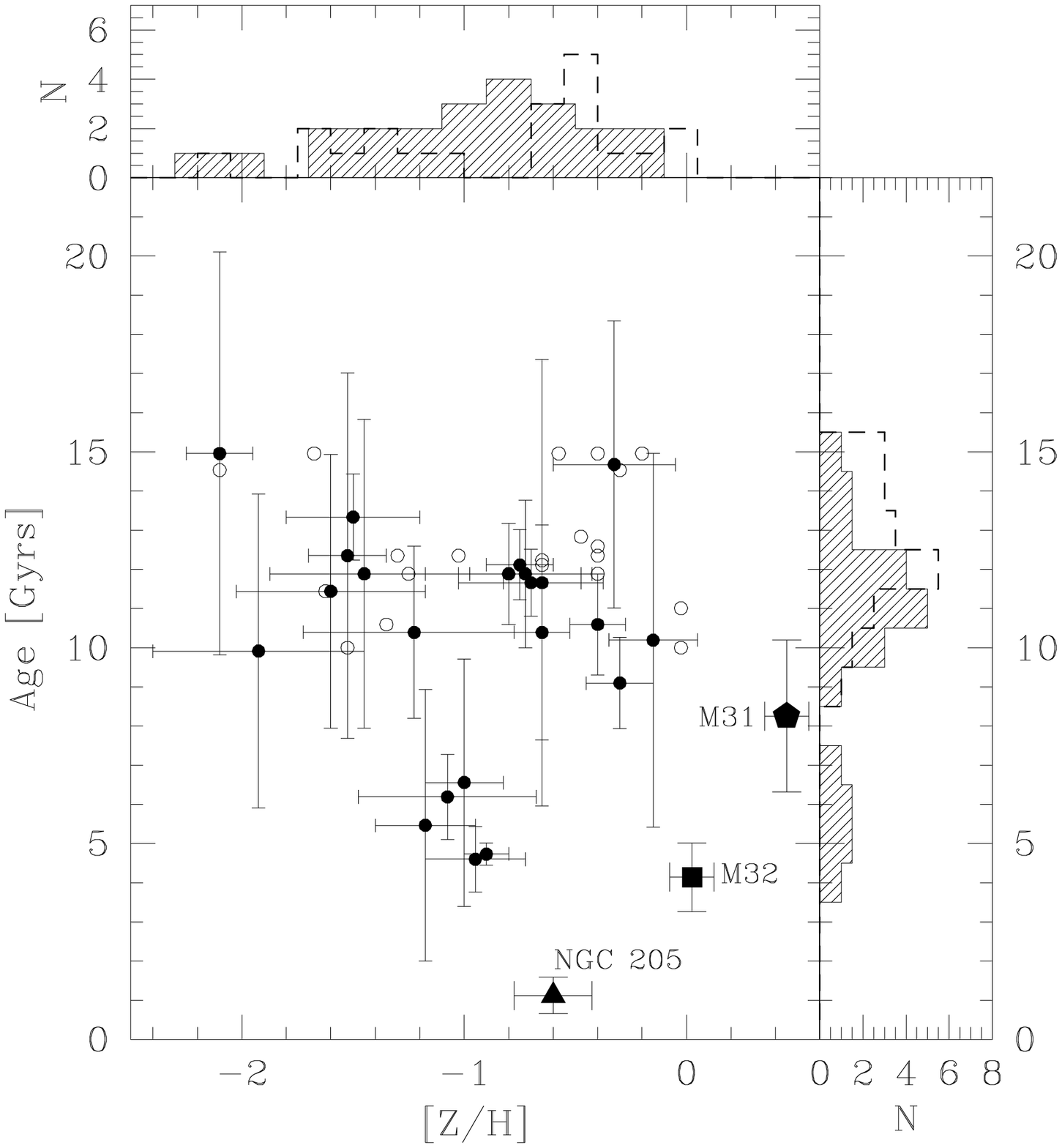}}
\figcaption{\small
The ages and metallicities of the M31 (filled circles with errors), 
and Galactic GCs (open circles) according to the Thomas, Maraston
\& Bender (2003) models. The shaded and open-dashed histograms 
exclusively represent  the M31 and Galactic GCs respectively.
Also shown are the positions of NGC~205, 
M32 and the bulge of M31, data taken from Trager et al. (1998).
\label{TMKresults}}
\end{center}}

\vbox{
\begin{center}
\leavevmode
\hbox{%
\epsfxsize=10cm
\epsffile{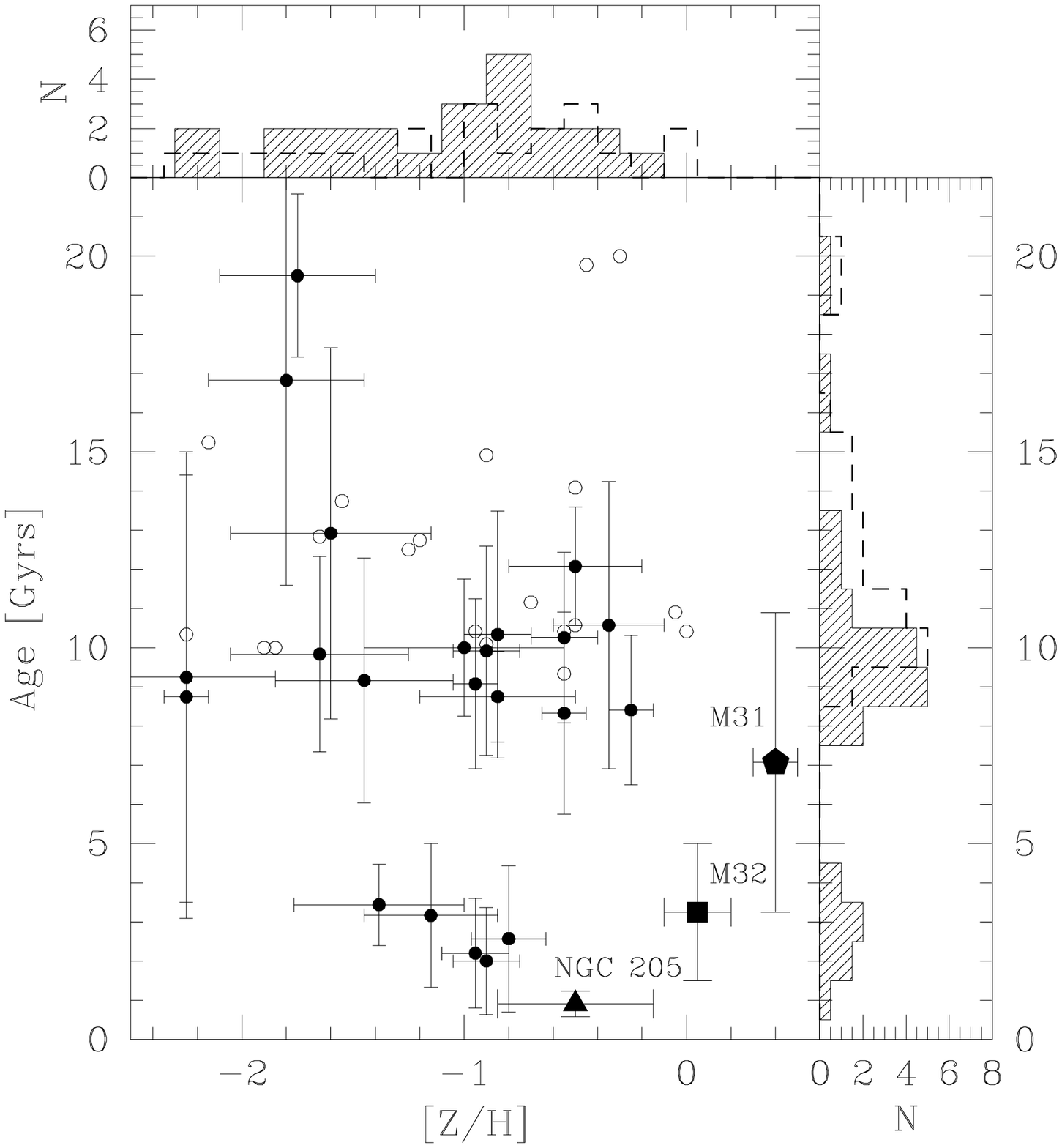}}
\figcaption{\small
The ages and metallicities of the M31 and Galactic GCs 
according to the Bruzual \& Charlot (2003) models.
Symbols as for previous figure.
\label{BC3results}}
\end{center}}

Our best-fit ages and metallicities for the
GCs are shown in Figures~\ref{TMKresults} and
\ref{BC3results}. Although in detail, the ages and metallicities
derived from the TMB03 and BC03 models differ, the basic 
conclusions remain the same. The M31 GCs separate out into 
three clear groupings: a metal-poor, old component (7 clusters), 
a metal-rich, old component (10 clusters) and 
an intermediate age, intermediate-metallicity component
(5 clusters). Reassuringly, all the Galactic GCs are 10 Gyr
old or older according to both sets of models.
As mentioned previously, the intermediate age GCs  
(henceforth IAGCs)
lie at around $\sim$5-6 Gyr according to the TMB03 models, 
whereas the BC03 models put them at 2-3 Gyr old.
Their ages and metallicities are not dissimilar to those
observed for star clusters in the Small Magellanic Cloud
(SMC; e.g., Da Costa \& Hatzidimitriou 1998).
Broadband optical colors are affected by age-metallicity
degeneracy and uncertain reddenings. However, the 
colors of the intermediate age candidates support
our spectroscopic findings (Section~\ref{IAGC}). Excluding
cluster NB~16, which possesses B--V=1.33 and we believe
is highly reddened, we obtain 
$\langle$(B--V)$_0\rangle$=0.7$\pm$0.1.
Assuming a metallicity of [Z/H]=--1.0, 
this corresponds to ages of roughly 5 Gyr
according to the BC03 and Maraston (1998) 
models.

The right-hand panels of Figures~\ref{TMKresults} and
\ref{BC3results} clearly show that the age distribution
in the M31 GCs is bimodal, whereas this is not the case 
for the Galactic GCs. Intriguingly, there is also
evidence that the mean ages of the old GCs
are different between the M31 and Galactic samples.
The old M31 GCs appear to be on average $\sim$2 Gyr
younger than the Galactic GCs, and this is driven
by the metal-rich ([Z/H]$>$--1.0) M31 clusters.
Since these are differential
comparisons, and are seen in the results for both models, this is
a fairly robust result. However, the analysis of a
larger, unbiased sample of clusters is required 
to check the reality of this result.
The metallicity distributions of the two GC samples 
appear quite similar, although the Galactic GC sample
extends to slightly higher metallicities. We identified
no super-solar metallicity M31 GCs in our sample.
One M31 GC in our sample, 225-280, possesses CMDs obtained
with HST/WFPC2 (Fusi-Pecci et al. 1996) and HST/NICMOS (Stephens et al. 2001). 
Our metallicity for this GC is TMB03 (BC03)
[Z/H]=--0.3$\pm$0.15 (--0.25$\pm$0.10). The two HST
studies are in good agreement, with [Z/H]=--0.4 and 
--0.15$\pm$0.37 respectively.

Also indicated in Figures~\ref{TMKresults} and
\ref{BC3results} are the positions of
NGC~205, M32 and the bulge of M31 (using the data
of Trager et al. 1998). The M31 bulge is clearly 
much more metal-rich than both GC samples, whereas M32
appears to have a metallicity comparable to the most
metal-rich Galactic bulge clusters NGC~6553 and NGC~6528.
Our ages and metallicities for both M31 and M32 are consistent
with previous findings (e.g. Mould, Kristian \& Da Costa 1984; 
Terlevich \& Forbes 2002; Schiavon et al. 2004a).
In terms of age and metallicity, the IAGCs are 
closer to that of the dwarf elliptical NGC~205.
Our solutions for NGC~205 indicate that it
is both slightly more metal-rich ([Z/H]=--0.8 (TMB03)), and 
somewhat younger than ($\sim$ 1 Gyr) the IAGCs. This seems to suggest 
either very recent or ongoing star formation in this galaxy 
(Bica et al. 1990; Lee 1996; Demers, Battinelli \& Letarte 2003).

On the CN abundances of these data, as discussed
in Section~\ref{Index}, the Milky Way and majority
of M31 clusters are strongly overabundant 
in the Lick CN indices when compared to SSP models
based on local stars. This is confirmed during the 
$\chi^2$ fitting procedure. The CN indices are 
rarely able to be fit when the best solutions for 
age, metallicity and abundance ratio are found
(see also PFB04). 
This is also true for the bulge of M31, which 
appears to be at least as overabundant in CN as the 
M31 and Galactic GCs. However, this is {\it not}
the case for any of the IAGCs, 
NGC~205 or M32 (see also Schiavon et al. 2004a). 
Since the CN index does not evolve
strongly with age, particularly after a few Gyr, 
this result implies that these objects do not
exhibit any significant CN-enhancement.

\vbox{
\begin{center}
\leavevmode
\hbox{%
\epsfxsize=12cm
\epsffile{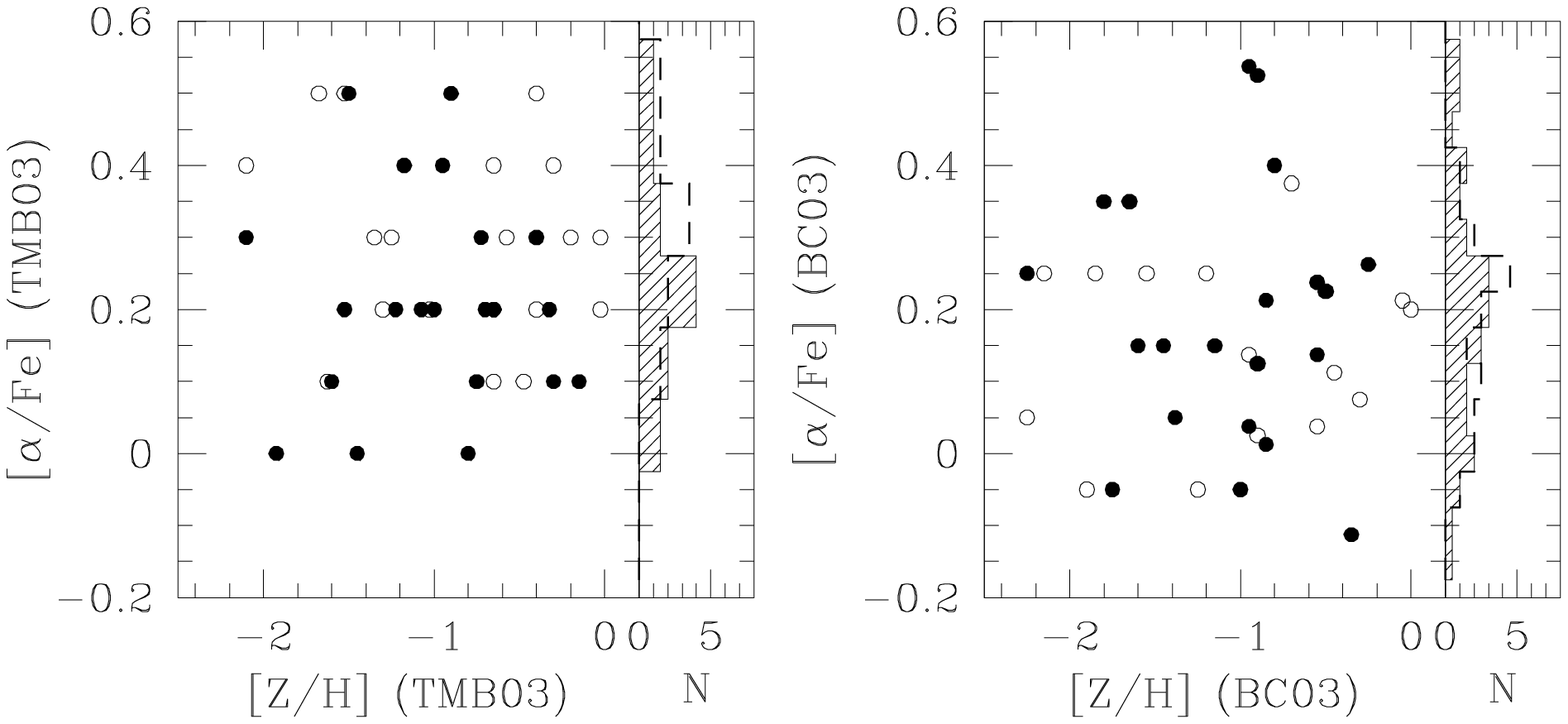}}
\figcaption{\small
The behavior of [$\alpha$/Fe] with [Z/H] for the 
M31 and Galactic GCs as inferred from the 
Thomas, Maraston \& Bender (2003) and
Bruzual \& Charlot (2003) models.
Symbols are as for the previous figure.
No significant trend is seen with metallicity.
\label{efe_ZH}}
\end{center}}

The behavior of [$\alpha$/Fe] with [Z/H] for the 
GCs is shown in Figure~\ref{efe_ZH}.
The vast majority of GCs exhibit [$\alpha$/Fe]$>$0, 
which is consistent with Figure~\ref{TMBMgbFe}.
There is no clear trend with metallicity; 
the Galactic and M31 GCs appear to be 
similarly enhanced at all values of [Z/H]. 
Interestingly, the TMB03 models suggest
that  the Galactic GCs are some $\sim$ 0.1 dex
more enhanced in [$\alpha$/Fe] than the M31
GCs (c.f. Figure~\ref{TMBMgbFe}).
However, this is not supported by the 
BC03 models which suggest similar levels 
of enhancement between the two populations.
Therefore, in this regard, it is difficult to come
to significant conclusions.
We have also determined [$\alpha$/Fe] ratios
using the TMB03 (BC03) models for NGC~205, M32 and M31 
and find [$\alpha$/Fe]= --0.3$\pm$0.1(--0.3$\pm$0.1), 
--0.1$\pm$0.1(0$\pm$0.1) and 0.2$\pm$0.1(0.1$\pm$0.1)
respectively. These values are consistent with previous
studies based upon integrated spectra 
(Terlevich \& Forbes 2002; Schiavon et al. 2004a).

\vbox{
\begin{center}
\leavevmode
\hbox{%
\epsfxsize=12cm
\epsffile{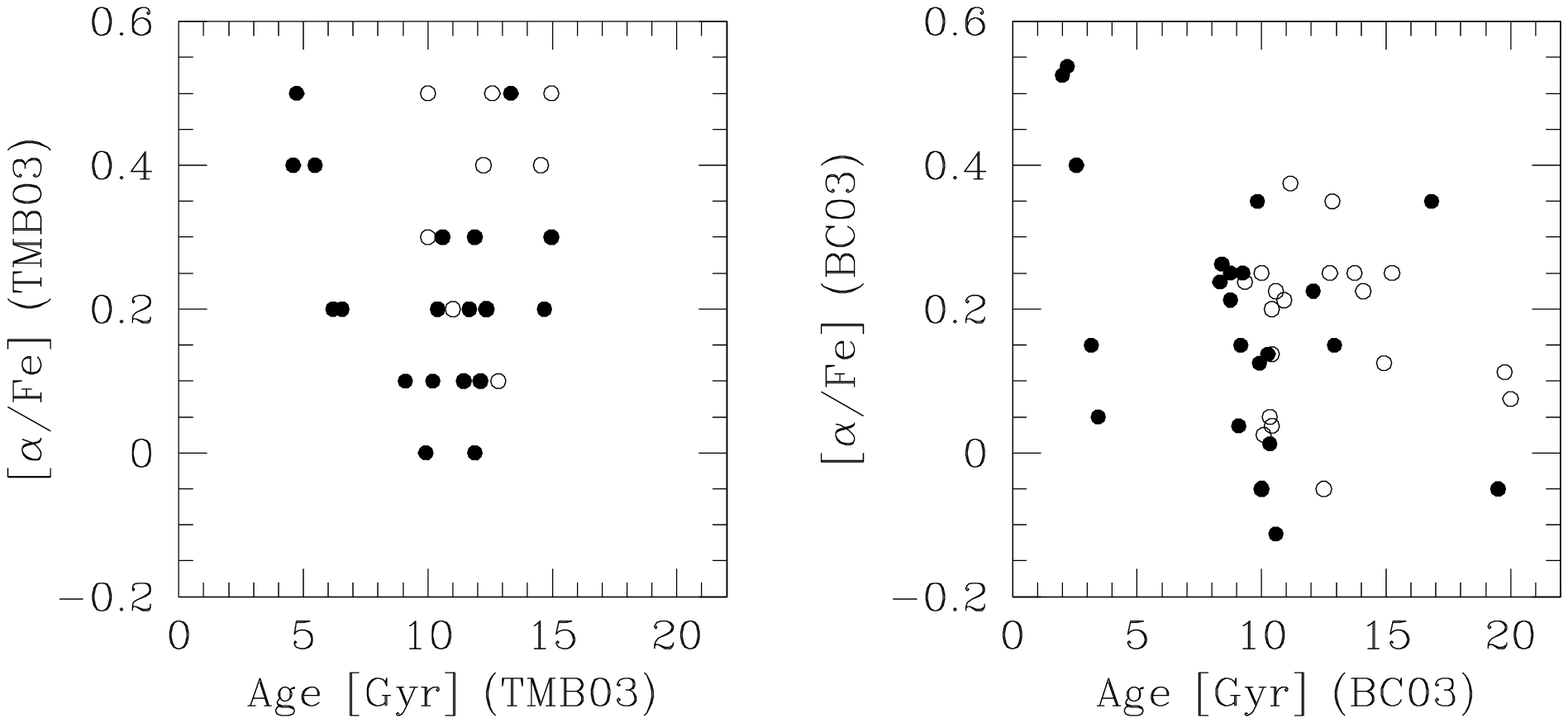}}
\figcaption{\small
The behavior of [$\alpha$/Fe] with age for the 
M31 and Galactic GCs as inferred from the 
Thomas, Maraston \& Bender (2003) and
Bruzual \& Charlot (2003) models.
Symbols are as for the previous figure.
The IAGCs appear to have
significant [$\alpha$/Fe] enhancement.
\label{efe_age}}
\end{center}}

Figure~\ref{efe_age} shows the behavior
of [$\alpha$/Fe] with GC age.
The IAGCs seem to have on average
higher [$\alpha$/Fe] ratios than the old
GCs using both models. A Kolmogorov-Smirnov test 
finds that this is significant at the 
2$\sigma$ level.
This may suggest that the material from which 
these objects formed was preferentially 
exposed to the ejecta from type-II supernovae, 
similar to the situation for young
clusters in the merger remnant NGC~3610
(Strader et al. 2004).
The $\alpha$-abundance pattern at [Z/H]$\sim$--1.0
is consistent with that seen for 
intermediate age Large Magellanic 
Cloud (LMC) clusters (R. Proctor, private comm.).
The TMB03 models suggest a slight correlation
between age and [$\alpha$/Fe] for the old M31 GCs, 
in the sense that the [$\alpha$/Fe] ratios are lower for 
younger clusters. This is, however, not apparent in 
BC03 model predictions.

In summary, we find that the M31 GCs in our sample 
separate into three distinct groups in age and metallicity.
We identify an old, metal-poor group, an old metal-rich group
and an intermediate age, intermediate metallicity group.
This result does not depend upon the SSP model employed.
The [$\alpha$/Fe] ratios of the M31 GCs are 
somewhat more model-dependent, but in general we find that the GCs 
have [$\alpha$/Fe]$>$0 similar to, perhaps slightly lower than, the 
Galactic GCs. The IAGCs appear strongly
enhanced in $\alpha$-elements.

\subsection{The Nature of the Intermediate-Aged Clusters}
\label{IAGC}

Previous spectroscopic studies of M31 GCs 
have indicated that photometric catalogs 
of M31 GCs can suffer from contamination
due to foreground stars (e.g., Barmby et al. 2000; 
Perrett et al. 2002). 
Kinematics of the M31 GCs does
not uniquely determine their identity, 
since the velocity fields of both the 
Galactic stellar disk 
and halo overlap with the M31 GC velocity 
distribution. For example, at the 
Galactic co-ordinates of M31 ($l$=121.17$^{\circ}$, 
$b$=--21.57$^{\circ}$) foreground disk F-dwarfs
are expected to have a mean streaming velocity 
of $\sim$ --62 kms$^{-1}$ according
to the Besan\c{c}on\footnote{bison.obs-besancon.fr/modele/} 
Galactic model (Robin et al. 2003).
In terms of visual identification, due to the
proximity of M31 (D$\sim$780 kpc, Holland 1998; Stanek \& Garnavich 1998),
GCs are only distinguishable  from stars from the ground in 
sub-arcsecond seeing conditions.
In view of this, it is important to ask whether
the IAGCs we have identified are truly 
clusters in M31, rather than foreground disk or halo stars.

Two of the IAGCs, 292-010 and 337-068, have been 
partially resolved by Racine (1991) using 
high-resolution CCD imaging. He identifies these
as {\it bona fide} clusters and we have no reason to doubt 
these identifications. The catalogs of Battistini et al. (1993,1987) 
and Sargent et al. (1977) have been shown 
to be partially contaminated with foreground stars
and galaxies (e.g. Barmby et al. 2000).
These photometric studies include the IAGCs 126-184, 301-022, 
NB16 and NB67. We searched for archival ground-based CCD imaging
with sub-arcsecond seeing and an adequate 
pixel scale, and found I-band imaging for IAGCS 126-184, 301-022
and NB16 in the NOAO NSA archives (M31F5-I and 
M31F9-I, PI Dr P. Massey)\footnote{archive.noao.edu/nsa}
taken with the Kitt Peak National Observatory 4-m
8K$\times$8K Mosaic-1 CCD. From measuring the point-spread
functions (PSFs) of several stars in the images we determined the
seeing to be 0.7 arcsec and 0.9 arcsec for
the F5-I and F9-I fields respectively.
We checked and verified that these PSFs did not vary significantly 
across the field of view of these data.

\vbox{
\begin{center}
\leavevmode
\hbox{%
\epsfxsize=10cm
\epsffile{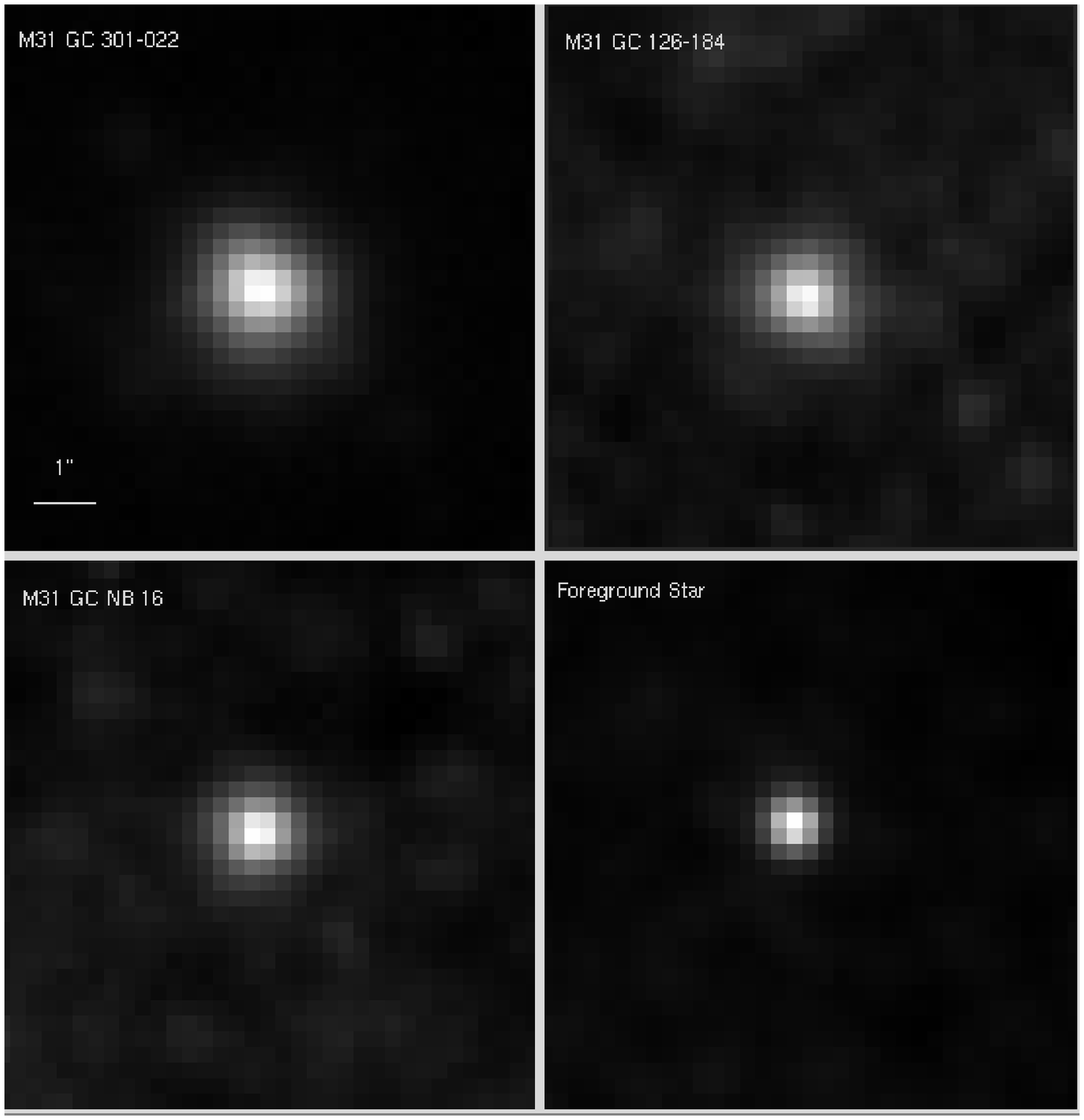}}
\figcaption{\small
KPNO Mosaic-1 I-band images of the IAGCs 301-022, 126-184
and NB16. The images are 9$\times$9 arcsec, and 
the pixel scale of the KPNO images is 0.258 arcsec/pixel.
Also shown is the appearance of a foreground
star identified from the USNO-A2.0 astrometric catalog, 
which has a similar I-band magnitude to the 
IAGCs. The same logarithmic image scaling and 
intensity cuts have been applied in each case.
The IAGCs are clearly less centrally concentrated
than the star.  
\label{globZ}}
\end{center}}

\vbox{
\begin{center}
\leavevmode
\hbox{%
\epsfxsize=10cm
\epsffile{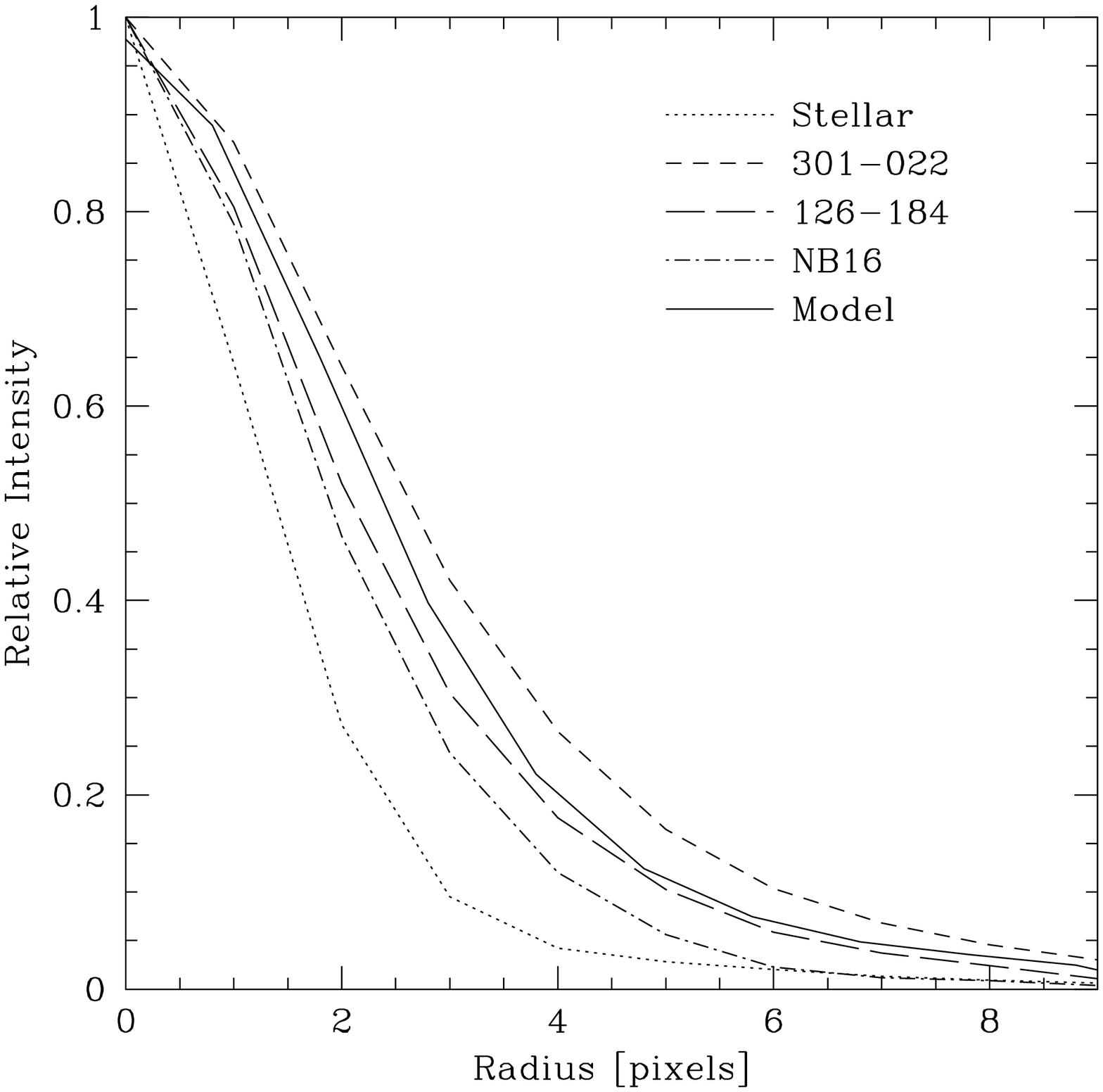}}
\figcaption{\small
Radial brightness profiles of the 
IAGCs 301-022, 126-184 and NB16 compared
to the comparison star shown in the previous
figure. All three IAGCs appear non-stellar.
Also shown is the profile of an artificial 
globular cluster at the distance of M31,
convolved with a Gaussian PSF (see text).
\label{profiles}}
\end{center}}

Postage-stamp images of the three IAGCs taken from
the KPNO frames are shown in Figure~\ref{globZ}.
The three clusters clearly appear more extended, and 
less centrally peaked, than the comparison star.
This is confirmed by the radial profiles of these objects 
which are shown in Figure~\ref{profiles}. In each case, 
the IAGCs show an excess of light beyond the 
PSF expected of a foreground star.
For comparison, we also show the brightness
profile of an artificial GC with a Hubble profile
(core radius of 0.2 pc), which has been convolved with a Gaussian
PSF of 0.7 arcsec and placed at the distance of M31. 
We conclude that these are non-stellar objects
with sizes consistent with compact globular clusters.

\vbox{
\begin{center}
\leavevmode
\hbox{%
\epsfxsize=10cm
\epsffile{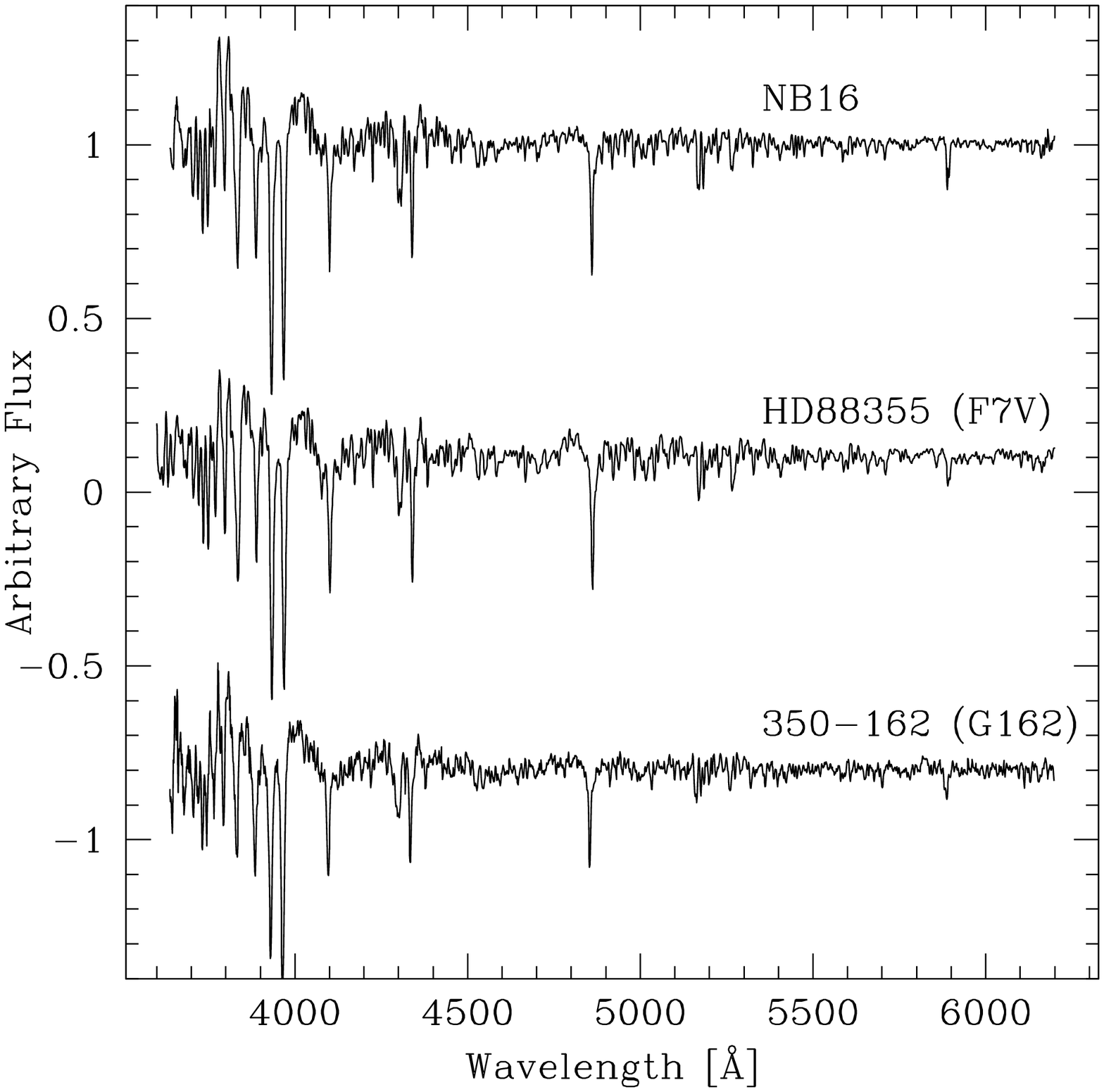}}
\figcaption{\small
Normalized spectra for a candidate IAGC (NB16; top), 
a Galactic F-dwarf from the STELIB library
(Le Borgne et al. 2003) and a true
metal-poor, old M31 GC (350-162; bottom).
\label{compare_spectra}}
\end{center}}

By cross-correlation with the STELIB
stellar library (Le Borgne et al. 2003)
and through visual inspection, we assign spectral
types to the IAGCs of F6--F7. Both strong 
Balmer lines and metallic lines are visible
characteristics of these spectral types.
In Figure~\ref{compare_spectra} we compare
the spectra of a candidate IAGC (NB16)
with that of a Galactic F-dwarf taken
from the STELIB library 
and an old, relatively metal-poor ([Z/H]$\sim$--1.5)
M31 GC (350-162). 
All three objects exhibit F-type spectra
and look quite similar. The 'inversion'
of the \ion{Ca}{2} H\&K lines indicates the presence of a 
hotter population
of stars in the spectrum of 350-162, which 
are presumably horizontal branch stars
(see Section~\ref{HBR}). 
NB16 and the Galactic F7 dwarf HD~88355
appear very similar; one difference
is the stronger strontium line ($\lambda$4077)
in the Galactic dwarf indicating metallicity
and/or luminosity differences.

Further, indirect evidence for the cluster nature
of the IAGCs comes from optical/near-IR colors, 
which potentially discriminate between warm stars 
and intermediate age stellar populations; 
the latter are expected to be redder 
in near-IR passbands for a given optical
color due to the flux contribution from
cooler stars on the giant branch 
(see e.g., Puzia 2003).

\vbox{
\begin{center}
\leavevmode
\hbox{%
\epsfxsize=10cm
\epsffile{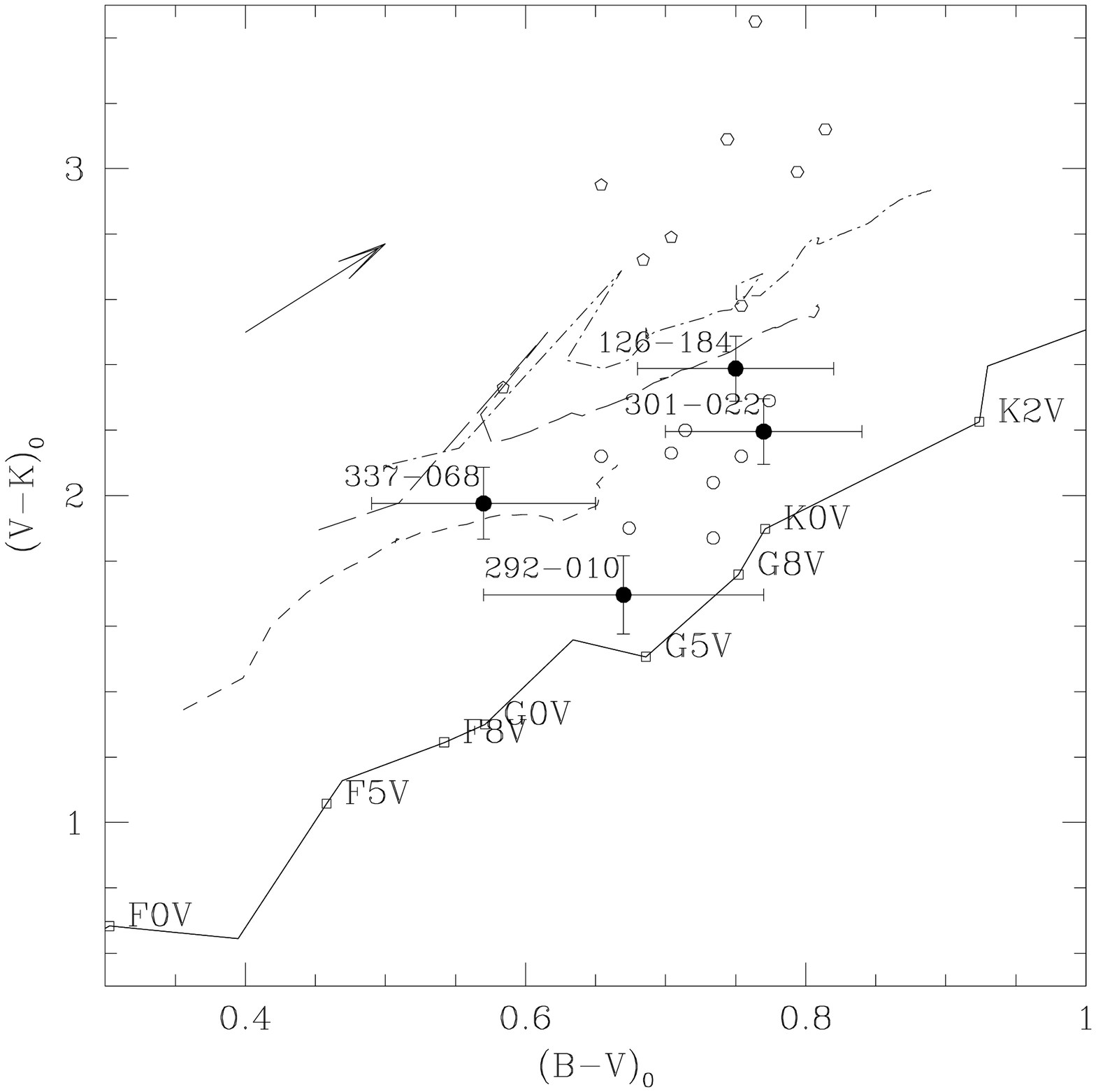}}
\figcaption{\small
Colors of four IAGC candidates (126-184, 
301-022, 337-068 \& 292-010) for which
exist optical and near-IR photometry.
Open symbols are for LMC clusters
(circles: SWB type VII; hexagons: SWB VI; 
pentagons: SWB V) taken from
Persson et al. (1983) and Bica et al. (1996).
Also shown is the stellar sequence
for F-K dwarfs in the solar neighborhood.
Short-dashed, long-dashed and dash-dotted
lines represent metallicities [Fe/H]= --1.65, --0.64 
and --0.33 respectively for ages 1--20 Gyr 
from Bruzual \& Charlot (2003).
The arrow corresponds to a reddening
of E(B--V)=0.1 (E(V--K)=0.27).
Three of the IAGCs lie clearly above
the stellar sequence. 
\label{colors}}
\end{center}}

The (B--V)$_0$,(V--K)$_0$ colors for four
IAGCs for which we have optical and near-IR
photometry are shown compared to the local stellar
sequence in Figure~\ref{colors}.
For clusters 126-184 and 337-068 we
do not have reliable reddening estimates
and therefore have assumed a mean E(B--V)=0.22
(Barmby et al. 2000).
Three IAGCs (126-184, 301-022 \& 337-068)
lie 0.3--0.5 mag above the stellar
locus; this cannot be attributable to reddening
since this vector runs parallel to this locus. 
The IAGC candidate 292-010 lies near the stellar locus, 
although as noted previously Racine (1991) identified this 
as a partially resolved object.
We also show in the figure the positions
of LMC clusters with SWB types V, V and VII
(corresponding to ages 0.8--2, 
2--5 and 5--16 Gyr respectively; Bica et al. 1992).
With the exception of 292-010, the IAGCs
lie somewhere between the old (SWB VII)
and younger (SWB V and VI) clusters, consistent
with our age estimates.

A schematic of how the M31 GCs in our sample are distributed
with respect to the optical disk of M31 is shown in Figure~\ref{spatial}. 
The old M31 GCs are widely distributed across the optical disk
and halo, and exhibit rotation 
in the sense of NE-approaching and SW-receding. This 
is consistent with the Perrett et al. (2002) results for a 
larger sample of GCs. Rotation in the young ($\leq$ 1 Gyr)
metal-rich ([Z/H]$\sim$0) disk clusters (Morrison et al. 2004; Paper I) 
is also apparent in Figure~\ref{spatial}.
Three of the IAGCs (126-184, NB16 and NB67)
are located near the bulge regions of M31, and
have comparable velocities (--182$\pm$14, --115$\pm$15
and --113$\pm$17 kms$^{-1}$ respectively).
The three other IAGCs (292-010, 301-022 and 337-068)
show no preferred spatial distribution, and have 
disparate velocities (--392$\pm$56, --30$\pm$30 
and 50$\pm$12 kms$^{-1}$ respectively).
The IAGCs show no signs of rotation but, 
excluding 292-010, do exhibit higher velocities 
than the bulk of the M31 GC system (see Figure~\ref{velocity}).

\vbox{
\begin{center}
\leavevmode
\hbox{%
\epsfxsize=10cm
\epsffile{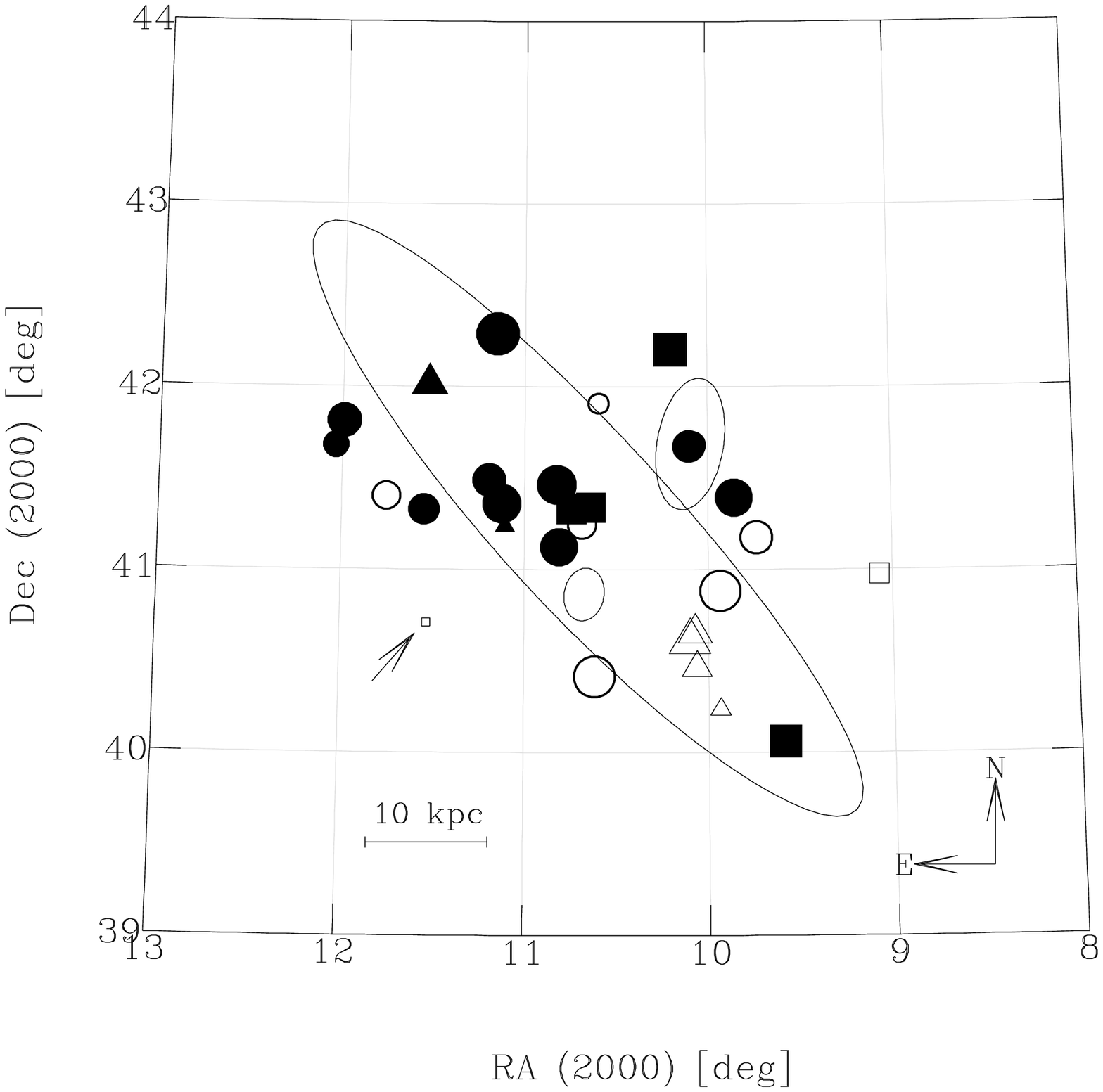}}
\figcaption{\small
Spatial distribution of M31 GCs, with respect to the M31 disk (large ellipse), 
NGC~205 (ellipse N of disk) and M32 (small ellipse).
Open symbols for old (open circles) and intermediate age
GCs (open squares) represent velocities lower than M31 
systemic velocity (--300 kms$^{-1}$). Filled symbols indicate velocities
higher than this, with symbol size proportional to the velocity 
residual from the systemic velocity. Note that three IAGCs are stacked
in the central regions, and have been offset slightly
for display purposes.
Also shown are the young disk clusters identified
in Paper I (open and filled triangles) and the ACS field of Brown et al (2003; 
indicated by an arrow).
\label{spatial}}
\end{center}}

The velocity distribution of the M31 sample is shown 
in Figure~\ref{velocity}. The mean 
velocity of our entire sample (including the disk 
clusters - Paper I) is --275$\pm$28 kms$^{-1}$, 
with a velocity dispersion of 155 kms$^{-1}$.
Excluding 292-010, the mean velocity 
of the IAGCs is --78$\pm$36 kms$^{-1}$, 
with a velocity dispersion of 80 kms$^{-1}$.
Monte Carlo tests by drawing velocities
randomly from the parent velocity distribution
indicate that this velocity difference
is statistically significant at the 99.8\% 
confidence level.
Also shown in Figure~\ref{velocity} is the velocity 
distribution for $\sim$200 clusters from Perrett et al. (2002).
Comparison between the velocity distribution of the full
M31 sample and that of Perrett et al. (2002) suggests
that we are probably missing a number of lower velocity GCs
($\sim$ --400 kms$^{-1}$). 

\vbox{
\begin{center}
\leavevmode
\hbox{%
\epsfxsize=10cm
\epsffile{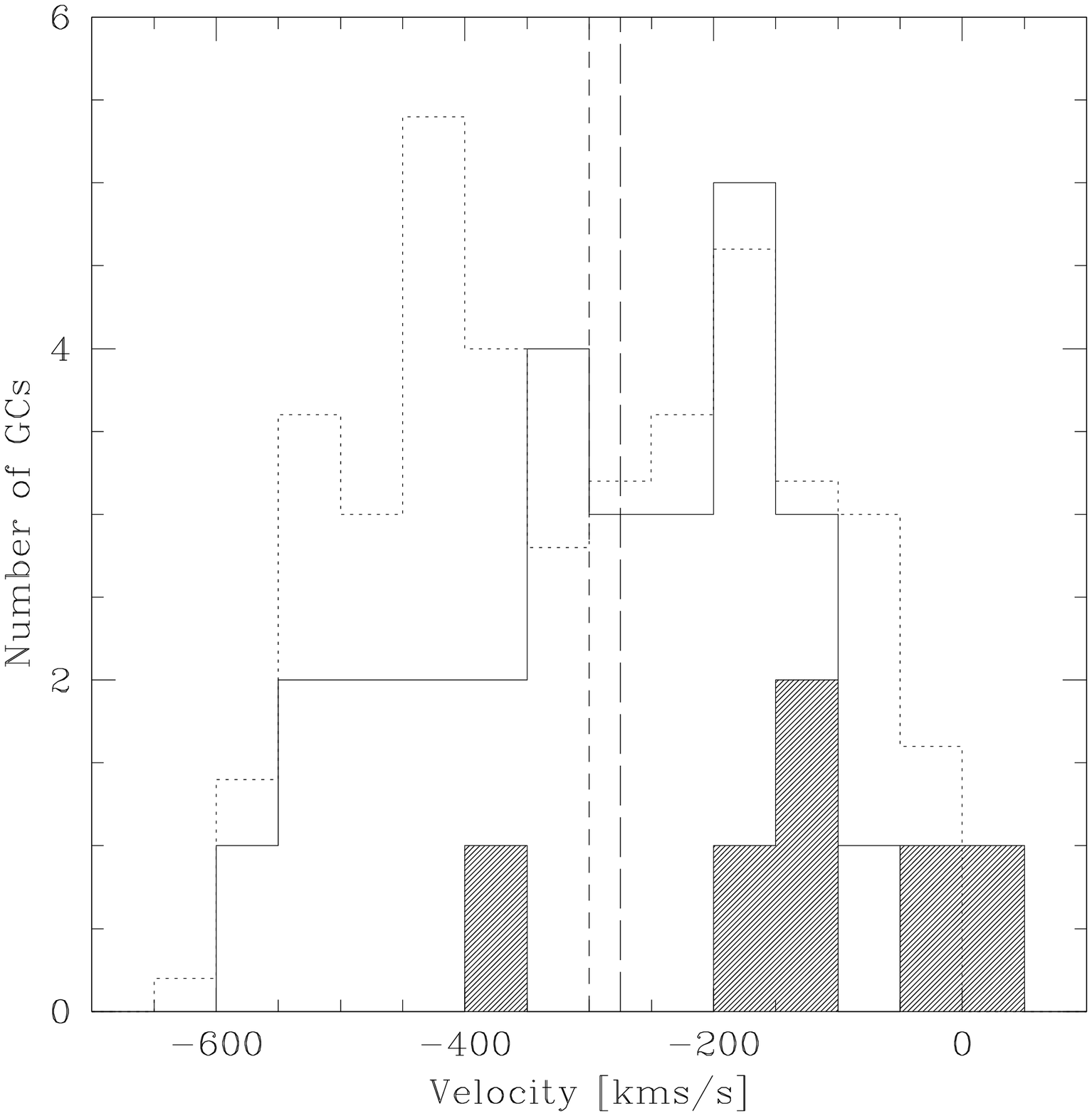}} 
\figcaption{\small
Velocity distribution of M31 GCs. The filled histogram 
indicates the velocities of the IAGCs, 
the open histogram is for our full sample.
The dotted histogram is for the sample of
$\sim$200 GCs from Perrett et al. (2002) scaled
down by a factor of 6. The vertical short-dashed line 
indicates the systemic velocity of M31 (--300 kms$^{-1}$), 
the vertical long-dashed line indicates the 
mean velocity (--275$\pm$28 kms$^{-1}$) 
of our full sample.
\label{velocity}}
\end{center}}

In summary, based upon the extended nature
and colors of the IAGCs, we conclude that 
clusters 126-184, 292-010, 301-022, 337-068
\& NB16 are star clusters 
associated with M31. At present we cannot
state confidently that NB67 is also a 
star cluster, although its similarity
to NB16 in terms of line-strengths
and kinematics makes this is a distinct 
possibility.

\section{The Horizontal Branch Morphologies of M31 GCs}
\label{HBR}

Age estimates based on Balmer lines can be seriously
affected by the presence of blue horizontal branches (HBs)
in GCs (Lee et al. 2000; Beasley et al. 2002; Peterson et al. 2003;
TMB03; Schiavon et al 2004b). Such stars have high temperatures, and 
can mimic the effect of hotter (younger) turn-off 
temperatures. The existence of varying HB morphology 
at a given metallicity in some GCs (the second parameter effect) further
complicates this issue (e.g., TMB03).
We can, however, rule out the possibility that hot HB stars
are conspiring to make the IAGCs look 
young in their integrated spectra through three arguments:

Firstly, the multivariate approach which we have employed 
is robust against such a possibility. This has been 
demonstrated by PFB04, and is essentially true because
reliable age estimates can be determined without the use
of Balmer lines in the age determination. This can be achieved
because each index retains some age and metallicity 
information. By using sufficient indices, the age-metallicity
degeneracy can be broken without the use of the age-sensitive
Balmer lines. Since these indices are the most
temperature sensitive (and hence HB morphology sensitive) 
Lick indices, omitting these from the $\chi^2$ fit
allows the potential effects of HB stars to be mitigated.

Secondly, Schiavon et al. (2004b) have 
shown that the age-HB morphology degeneracy can be broken
by comparing the strengths of the Balmer lines.
They argue that bluer Balmer lines should be effected more 
strongly by the presence of blue HB stars because 
at the characteristic temperatures of such stars ($\sim$ 9000K), 
their relative flux contribution increases for bluer indices.
Thus, H$\delta$, H$\gamma$ should be relatively stronger 
than H$\beta$ in the integrated spectrum of an HB-affected 
GC, when compared to a canonical model grid.
Inspection of Figures~\ref{hbeta}, \ref{hgamma_a} and
\ref{hdelta_f} indicates 
that this is not the case here, in fact the H$\beta$ index
appears much stronger in the IAGCs
relative to the older M31 clusters than do the higher-order
Balmer lines. Conversely, this difference in
separation between H$\beta$ and the higher-order
Balmer lines could be qualitatively explained
if one assumes that the IAGCs have purely red HBs, 
and the Balmer lines of the 'old'
metal-poor GCs are affected by HBs. However, then the only explanation
for the position of the IAGCs in the H$\beta$--[MgFe] plane
(Figure~\ref{hbeta}) is to invoke intermediate ages.

Thirdly, we may estimate the HB morphology of the M31 GCs
by using the Rose (1984) \ion{Ca}{2} diagnostic (see also Tripicco 1989; PFB04).
The \ion{Ca}{2} index, defined as the ratio of the 
\ion{Ca}{2} H+H$\epsilon$ to \ion{Ca}{2} K line, 
is sensitive to the presence of hot stars in evolved stellar populations (Rose 1985).
This index is extremely effective in identifying hot stellar
populations because it is a ratio of two strong lines
of singly ionized calcium, which implicitly arise from the
same physical process. Since H$\epsilon$ is coincident
(in wavelength) with the \ion{Ca}{2} H line, hot stellar populations
will strengthen this line without effecting \ion{Ca}{2} K.
As shown by PFB04, this index shows a remarkable correlation with 
horizontal branch (HB)
morphology; \ion{Ca}{2} increases as HBs become bluer due to the 
strengthening of \ion{Ca}{2} H+H$\epsilon$ relative to \ion{Ca}{2} K.

\vbox{
\begin{center}
\leavevmode
\hbox{%
\epsfxsize=10cm
\epsffile{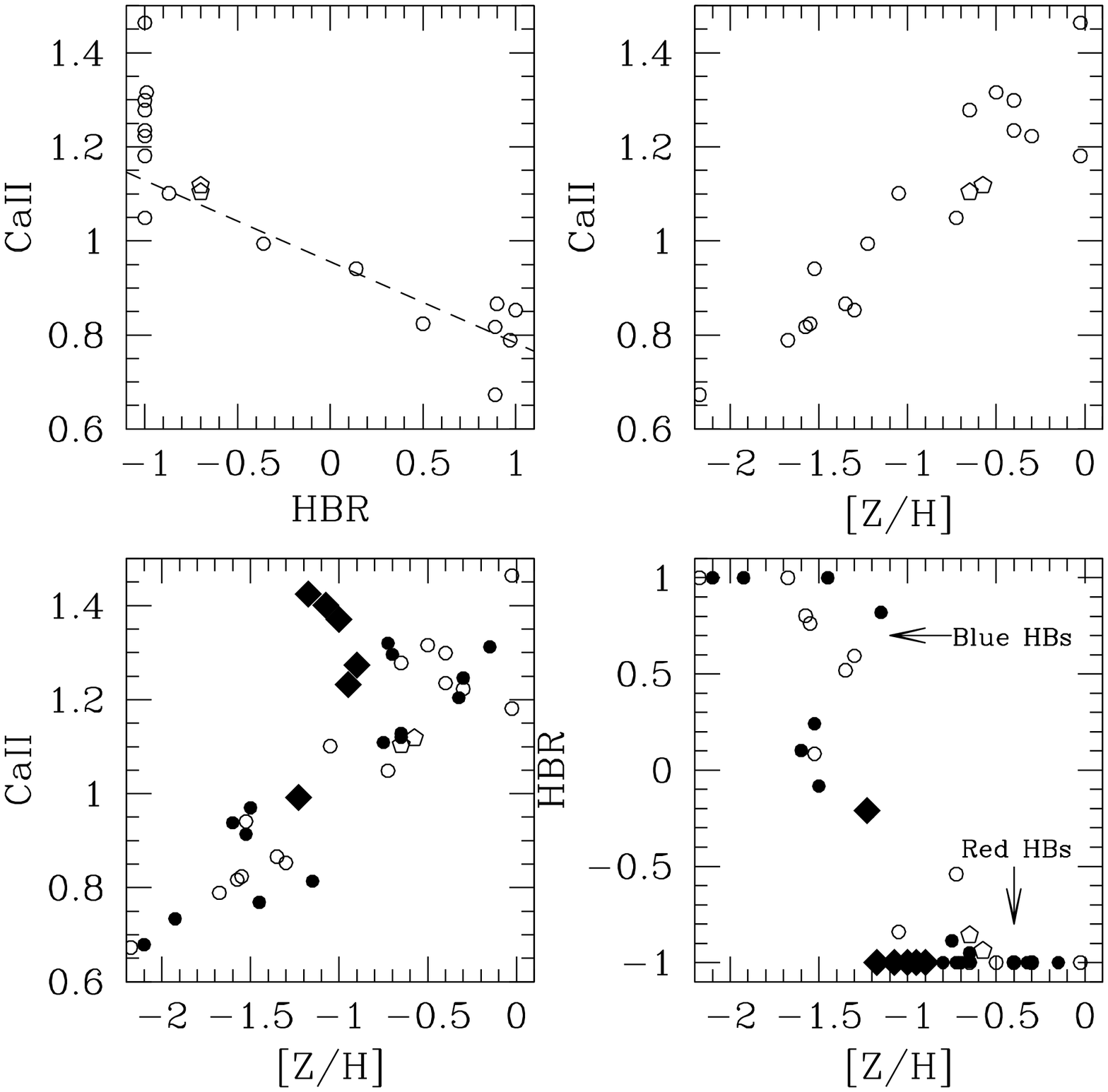}}
\figcaption{\small
Calibration of the Rose (1984) \ion{Ca}{2} index for 
horizontal branch morphology.
The top-left panel shows the good correlation between
the horizontal branch parameter and the \ion{Ca}{2} index
for Galactic GCs (open symbols). The two open
pentagons represent the Galactic GCs 
NGC~6388 and NGC~6441, which are metal-rich but 
are known to contain a blue HB component (Rich et al. 1997).
The dashed line is a fit to these data (see text).
The large solid diamonds indicate the 
IAGCs.
The top-right panel shows
the correlation between \ion{Ca}{2} and our derived 
metallicities for the Galactic GCs using the 
Thomas, Maraston \& Bender (2003) models.
The bottom-left panel shows the same
correlation, but includes the M31 GCs (solid
symbols). 
The bottom-right panel shows our derived HB morphology
for all the GCs. 
\label{HBRplot}}
\end{center}}

Figure~\ref{HBRplot} shows the calibration of the HB parameter 
(Lee et al. 1994) for the Milky Way and M31 GCs.
We have measured the \ion{Ca}{2} index for the P02 and M31 data, the
former which we have augmented with data for seven additional
Galactic GCs (47 Tuc, NGC~362, NGC~1466, NGC~1851, NGC~1904, 
NGC~6652 and NGC~7099) described in Gregg (1994). Seven clusters
are in common between the P02 and full Gregg (1994) sample, and we find 
excellent agreement; (P02-Gregg)=0.012, $\sigma$=0.062 in 
\ion{Ca}{2} between the two.
The HB parameter is taken from the Harris (1996) catalog in 
all but one case; NGC~1466 in the Gregg (1994) sample.
This is actually an LMC cluster for which we adopt the
HB ratio in Johnson et al. (1999).
The top-left panel in Figure~\ref{HBRplot}
illustrates the good correspondence between \ion{Ca}{2} and the HB parameter.
Unfortunately the HB parameter saturates before \ion{Ca}{2}, however, blue
HBs can clearly be separated from red HBs, while intermediate
HB morphologies may be estimated.

We calibrate the \ion{Ca}{2} index to indicate HB morphology as:

\[ {\rm HBR} = \left\{\begin{array}{ll}
	+1, 		& \mbox{\ion{Ca}{2}$\le$0.6}\\
	-4.99\times(\mbox{\ion{Ca}{2}})+4.77, & \mbox{0.6$<$\ion{Ca}{2}$<$1.2}\\
	-1,		& \mbox{\ion{Ca}{2}$\ge$1.2}
	\end{array}
	\right. \]

\noindent where HBR=--1 is equivalent to purely red HBs, +1 
equivalent to purely blue HBs and intermediate values corresponding 
to intermediate HB morphologies.
Our measured values for \ion{Ca}{2}, and the corresponding values for
the HB parameter for the M31 GCs are listed in Table~\ref{Solutions}.

Figure~\ref{HBRplot} (bottom-left) compares the 
the \ion{Ca}{2} index with our derived (TMB03) metallicities
for the Galactic and M31 GCs. The correlation is good, 
with the exception that the IAGCs which generally have
too high \ion{Ca}{2} for their metallicities.
The bottom-right panel shows the HB ratio, derived from 
\ion{Ca}{2}, versus (TMB03) metallicity. 
We find using the BC03 metallicities yields 
similar results in this calibration.
The M31 GCs clearly 
possess a range of HB morphologies, from 
very blue, to intermediate morphologies, to very red.
The majority appear to possess purely red HBs, 
which is consistent with recent GALEX UV measurements 
(M. Rich, private comm.). All but one of the IAGCs
are included in this category. IAGC 292-010
shows evidence for an 'intermediate' HB morphology.

Two Galactic GCs which are metal-rich, but exhibit 
blue HB components (Rich et al. 1997), are indicated
in Figure~\ref{HBRplot}. These clusters lift slightly from the 
'pure red HB' value of HBR=--1.
This suggests that the \ion{Ca}{2} method
is sensitive to such populations. 
Two M31 GCs also show similar behavior, 158-213 and 234-290, 
and are candidate metal-rich GCs in M31 which 
may have blue HB components.
Clearly this latter effect only weakly registers 
in \ion{Ca}{2}, and therefore relatively reliable 
metallicity estimates are required for it to be 
successful. 

Based upon the previous arguments, we conclude that 
it is unlikely that blue HBs are having a significant 
impact upon our age (or metallicity) determinations, and
that the IAGCs possess intermediate ages, rather than 
old ages and blue horizontal branches.

\section{Discussion}
\label{Summary}

Using the stellar population models of Bruzual \& Charlot (2003)
and Thomas, Maraston \& Bender (2003) we have estimated 
metallicities, ages and [$\alpha$/Fe] ratios for 
21 globular clusters in the Milky Way and 23 
globular clusters in M31. Our principle 
findings are as follows:

\begin{itemize}

\item as expected, we find that the Milky Way clusters are
uniformly old (within our measurement uncertainties), 
span a metallicity range,
--2.0$\leq$[Z/H]$\leq$0, and have mean 
[$\alpha$/Fe] ratios of 0.2$\sim$0.4 (depending 
upon the models employed).

\item the M31 clusters span a very similar metallicity 
range to the Milky Way clusters, and are also
enhanced in $\alpha$-elements.
There is some evidence that the old, metal-rich 
([Z/H]$>$--1.0) clusters in M31 are $\sim$ 2 Gyr
younger than the equivalent Milky Way GCs, 
although a larger sample of M31 GCs will test the 
validity of this result.

\item there is weak evidence that the M31 clusters 
have lower [$\alpha$/Fe] ratios than their Milky Way
counterparts. We see no evidence
for a correlation between [Z/H] and [$\alpha$/Fe].

\item M31 has a population of intermediate age (2--6 Gyr), 
intermediate-metallicity ([Z/H]$\sim$--1.0) globular
clusters (IAGCs). Through consideration of 
high-resolution CCD imaging and optical/near-IR colors for 
a subset, we show that it is unlikely 
these objects are interloping foreground stars.

\item stellar population models suggest that these  IAGCs
exhibit $\langle$[$\alpha$/Fe]$\rangle\sim$0.3--0.5, 
and, with the exception of one cluster identified
in Paper I, possess higher velocities than the bulk of the M31 cluster sample.
Three of the IAGCs are concentrated in the bulge regions of 
M31, the remaining three lie in the 'halo' 
regions. 

\item when compared to the rest of the M31
and Milky Way sample, the IAGCs have low CN values.
Furthermore, the IAGCs appear to be 'CN-depressed'
with respect to stellar population models.

\end{itemize}

Although our sample of M31 GCs is not unbiased, 
the implications of the
relative numbers of cluster sub-populations identified
in our spectroscopic sample are surprising. 
Including the six young disk clusters
identified in Paper I (and excluding Hubble V 
in NGC~205), from a total sample 
of 29 M31 globular clusters: 17 ($\sim$58\%) are old (metal-poor
and metal-rich), 6($\sim$21\%) are intermediate age and 
intermediate-metallicity and 6($\sim$21\%) are metal-rich, 
young disk clusters. van den Bergh (1999) estimated that the total
population of M31 GCs is 400$\pm$55, with 337 confirmed 
GCs (Galleti et al. 2004). 
Using the estimates of van den Bergh, 232$\pm$32 GCs in M31
are old (metal-poor and metal-rich), and there are 
84$\pm$12 intermediate age globular clusters, with a similar
number of young disk clusters.
This is highly likely to be a significant overestimate, 
and a complete spectroscopic sample of M31 GCs is required in order to check 
these crude values.

It is, at present, unclear whether the disk 
clusters are truly globular clusters, or simply represent the 
upper end of the open cluster luminosity function (Paper I).
However, the IAGCs have luminosities
(and hence masses, accounting for age-fading)
consistent with normal globular clusters 
(15.3$\leq V_0\leq$17.5; 4$\times10^4\leq L_{\rm GC}
\leq5\times10^5\Lsun$).
The spatial distribution of the IAGCs hints that any 
globular cluster formation which may have occurred 2--6 Gyr ago
in M31 was not confined to the bulge regions of this 
Galaxy. This suggests that the formation of the IAGCs 
cannot be explained by secular evolution in
the inner-disk/bulge regions (e.g., Maoz et al. 2001), 
without some mechanism for transporting a significant
fraction out to large radii.  

Interestingly, although unfortunately not included in the 
present sample, there is evidence
to suggest that several of the outer halo Milky Way  GCs
may be younger than their inner-halo counterparts.
For example, color-magnitude diagrams have suggested that 
the outer halo GCs Pal 12 (Gratton \& Ortolani 1988; Stetson et al. 1989; 
Rosenberg et al. 1998) and Ruprecht 106 (Buonanno et al. 1990; 
Da Costa et al. 1992; Buonanno et al. 1993) may have ages up to 
30\% younger than old halo GCs (i.e., as young as $\sim$ 8 Gyr).  
It has been suggested that Pal 12 could have 
been captured from the Sagittarius dwarf spheroidal (Sgr dSph) (Irwin 1999), 
and this notion is supported by the fact that Pal 12 lies among the tidal 
debris presently being stripped from the Sgr dSph (Martinez-Delgado et al.
2002; Bellazzini et al. 2003). Further evidence for this idea 
comes from the $\alpha$-capture elements 
of Pal 12 (Brown et al. 1997; Cohen 2004) and Rup 106 (Brown et al. 1997), which  
do not appear to be enhanced with respect to the solar values, but are consistent 
with the abundances of Sgr dSph stars (Bonifacio et al. 2000; 
Smecker-Hane \& McWilliam 2002). It is possible, although at present speculative, 
that the intermediate-aged clusters we have identified in M31 are younger 
analogs of these Milky Way GCs.

Observations of stellar substructure in M31
(e.g., Ibata et al. 2001; McConnachie et al. 2004),
suggest that mergers and/or accretions have played an 
important role in the growth of M31's halo
stellar populations.
One possibility is that M31 underwent a major 
merger $\sim$6 Gyr ago ($z\sim0.6$ for a standard flat cosmology), 
forming GCs and field stars in the process.
Indeed, Brown et al. (2003) have presented evidence
for an intermediate age (6--8 Gyr) population of stars
in the halo of M31, which they argue may hint at such an event. 
During such a merger, any pre-existing metal-rich
stellar component (e.g., bulge or thick-disk stars) is expected 
to be partly redistributed to large radii as observed in M31's
halo. The putative thin-disk in M31 (Morrison et al. 2004)
is not a strong constraint on the nature 
of the hypothetical merger, since the disk may in fact be
rather young ($\sim$ 1 Gyr, Paper I). 
A $\sim$ 6 Gyr hiatus is potentially sufficient time for 
shocked gas to cool into a new disk configuration.

However, the relatively low metallicities of the IAGCs
makes a major merger origin 
look somewhat unlikely, unless the merger progenitors were 
surprisingly metal-poor. The merging of 
two present-day L$_*$ spirals is expected 
to produce rather metal-rich ([Z/H]$\geq$0)
field stars and star clusters (Bekki et al. 2002).
Such major merging has to occur relatively early
in order to reproduce the red peak of 
elliptical galaxy GC systems (Beasley et al. 2002).
An additional constraint is that the [$\alpha$/Fe] 
ratios of the IAGCs imply that the proto-IAGC clouds 
were preferentially enriched by Type-II
supernovae. This abundance pattern is unexpected based on Milky Way disk
abundances (e.g., Edvardsson et al. 1993).
Perhaps less radical minor merger(s) or accretion(s) also provide
a route to forming star clusters under the proviso that 
sufficient gas is available.
The mass of the satellite, its density profile and 
orbital parameters would determine the degree of heating
of any disk present, but would not necessarily lead to its
disruption (Huang \& Carlberg 1997; Vel$\acute{a}$zquez \& White
1999).

We can derive an approximate upper limit on the 
neutral gas mass of the infalling satellite from the total 
luminosity of the six IAGCs ($\sim 6\times10^5\Lsun$). 
Assuming a mass-to-light ratio of 1 for the IAGCs (BC03), and a
formation efficiency (i.e., M$_{\rm GC}$/$M_{\rm gas}$) 
of 0.25\% (McLaughlin 1999), 
we obtain $\sim2.4\times10^8\Msun$ of gas required.
Extrapolating the amount of gas necessary to form 
the estimated 84$\pm$12 IAGCs yields a required
\ion{H}{1} mass of $\sim3.4\times10^9\Msun$.
By comparison, the total \ion{H}{1} mass of the SMC
is 4.2$\times10^8\Msun$ (Stanimirovic et al. 1999), and 
that of the LMC is 5$\times10^8\Msun$ (Kim et al. 1998).
The very similar ages and metallicities of the 
IAGCs suggests that these objects formed
at a similar epoch, and from gas of a similar
metallicity. This in itself argues for a single 
progenitor, and for the 
clusters to be formed during the 
hypothetical progenitor's infall
onto M31. For example, the accretion of an
SMC-type satellite, and the subsequent accretion
of a pre-formed SMC-like cluster system, would yield
an age-metallicity relation inconsistent with 
that which we observe (e.g., Da Costa \& Hatzimiditrou
1998; Harris \& Zaritsky 2004).

It is difficult to try and associate (spatially
or kinematically) the IAGCs with the observed substructure in 
M31 without having detailed knowledge of the orbit of the satellite
giving rise to particular the substructure in
question. However, here we speculate briefly.
The preliminary orbit of the progenitor of
the 'Andromeda Stream' (Ibata et al. 2001)
appears to be surprisingly radial (and perhaps nearly co-planar) 
and passes close to the center of M31's bulge (Ferguson et al. 2002; 
Ibata et al. 2004). This is consistent with 
the location of three of the IAGCs. The orbit
to the North of the M31 bulge is less well constrained, but
would have difficulty passing through the location
of all the non-bulge IAGCs (c.f. Figure~\ref{spatial}).
Kinematically, the Andromeda Stream is at 
much lower velocities than five of the six IAGCs
(the mean velocity of the stream is $\sim$--400 kms$^{-1}$; 
Ibata et al. 2004), making a direct connection look
unlikely.

Ferguson et al. (2002) discuss the possibility that this
stream may be associated with M32. Could the IAGCs 
be associated with M32?
There is evidence that M32 has lost a large fraction
of its mass (Faber 1973), and has undergone tidal
interactions (e.g., Choi et al. 2002).
M32 is observed to have no GCs, although 
its luminosity would predict $\sim$15. These
have possibly been stripped by the tidal field
of M31 (van den Bergh 2000).
It has also been suggested that M32 once possessed 
significant quantities of gas, and is possibly
the remnants of a 'threshed' low-luminosity 
spiral (Bekki et al. 2001; but see
also Graham 2002).
This suggestion is intriguing since such 
a spiral may be expected to have a relatively
large reservoir of low-metallicity gas.
Based upon its integrated spectrum, M32
also has a central mean age similar to the IAGCs, 
and shows similar low CN values. However
it is also significantly more metal-rich 
([Z/H]$\sim$0) than the IAGCs, but this 
may simply reflect an extended period 
of star formation in this galaxy (e.g., 
Schiavon et al. 2004).

Another candidate for the progenitor
of the IAGCs is the dwarf elliptical 
NGC~205. In terms of its
light-weighted age and metallicity, this
galaxy is very similar to the mean properties
of the IAGCs (Mould, Kristian \& Da Costa 1984; 
Lee et al. 1996; Section~\ref{Multivariate}).
Similar to the case for M32 and the IAGCs, but unlike 
the bulge of M31, this galaxy exhibits 
low CN values.
It also shows signs of recent star formation
(Lee 1996), and the report of an 
arc-like overdensity spatially coincident
with NGC~205 suggests that this satellite is
undergoing some tidal disruption 
(McConnachie et al. 2004). 
However, without 
a knowledge of the orbit of this galaxy 
we refrain from speculating further on its
possible connection to the IAGCs.

In summary, we find that the Andromeda spiral appears
to have at least four sub-populations
of star clusters: one metal-poor and old, one metal-rich and
old, one intermediate age and of intermediate-metallicity,
and one very young with solar metallicity (Paper I).
These younger populations appear
kinematically distinct from their older
counterparts. A complete spectroscopic 
survey is necessary to help understand why the 
globular cluster system of M31 differs so much
from our own Galaxy.

\section{Acknowledgments}

We thank Michael Gregg for supplying his 
Galactic GC spectra in digital form, Javier Cenarro and 
Kenji Bekki for useful discussions.
This research has made use of the SIMBAD database,
operated at CDS, Strasbourg, France.
This research draws upon data provided by [Dr P. Massey] as distributed by the 
NOAO Science Archive. NOAO is operated by the Association of Universities for 
Research in Astronomy (AURA), Inc. under a cooperative agreement with the National 
Science Foundation.
Funding support comes from NSF grant AST 0206139.
The data presented herein were obtained at the
W.M. Keck Observatory, which is operated as a scientific partnership among
the California Institute of Technology, the University of California and
the National Aeronautics and Space Administration.  The Observatory was
made possible by the generous financial support of the W.M. Keck
Foundation. This research has made use of the NASA/IPAC Extragalactic
Database  (NED), which is operated by the Jet Propulsion Laboratory,
Caltech, under contract with the National Aeronautics and Space
Administration.

\clearpage

\begin{deluxetable}{lccccccccc}
\renewcommand{\arraystretch}{.6} 
%\rotate
\footnotesize
\tablecaption{Metallicities, Ages, Abundance Ratios and Horizontal Branch 
Morphologies of the M31 Clusters. \label{Solutions}}
\tablewidth{0pt}
\tablehead{
\colhead{Name} & \colhead{[Z/H]$^a$}   & \colhead{Age$^a$}   &
\colhead{[$\alpha$/Fe]$^a$} &
\colhead{[Z/H]$^b$}  & \colhead{Age$^b$} & \colhead{[$\alpha$/Fe]$^b$} & 
\colhead{\ion{Ca}{2}} & \colhead{HBR$^c$} &\colhead{CN Strong?} \\
\colhead{} & \colhead{(dex)}   & \colhead{(Gyr)}   &
\colhead{(dex)} &
\colhead{(dex)}  & \colhead{(Gyr)} & \colhead{(dex)} & 
\colhead{} & & 
}
\startdata
126-184 & --1.38&   3.4 & 0.05 &--1.18 &   5.5 &  0.40 & 1.424 & --1.00 & N\\	
$\pm$ &  0.38 &   1.0 &  0.20 &  0.22 &   3.5 &  0.10 & & &\\
134-190 & --0.95&   9.1 & 0.04 &--0.72 &  11.9 &  0.30 & 1.320 & --1.00 & Y\\	
... &  0.10 &   2.2 &  0.50 &  0.25 &   1.9 &  0.30 & & &\\
158-213 & --0.90&   9.9 & 0.12 &--0.75 &  12.1 &  0.10 & 1.109 & --0.89 & Y\\
...&  0.15 &   2.7 &  0.20 &  0.15 &   0.9 &  0.10 & & &\\
163-217 & --0.35&  10.6 &--0.11 &--0.15 &  10.2 &  0.10 & 1.312 & --1.00 & Y\\
... &  0.25 &   3.7 &  0.10 &  0.20 &   4.8 &  0.10 & & &\\
225-280 & --0.25&   8.4 & 0.26 &--0.30 &   9.1 &  0.10 & 1.246 & --1.00 & Y\\
...&  0.10 &   1.9 &  0.10 &  0.15 &   1.2 &  0.10 & & &\\
234-290 & --0.85&  10.3 & 0.01 &--0.65 &  11.7 &  0.20 & 1.120 & --0.95 & Y\\
...&  0.15 &   3.2 &  0.10 &  0.17 &   5.7 &  0.10 & & &\\
292-010 & --1.35&  2.7 & 0.60 & --1.23 & 5.9 & 0.50 & 0.992 & --0.21 & N \\
...&  0.55 &   1.2 &  0.20 &  0.18 &   3.4 &  0.10 & & &\\ 
301-022 & --1.15&   3.2 & 0.15 &--1.08 &   6.2 &  0.20 & 1.501 & --1.00 & N\\
...&  0.30 &   1.8 &  0.10 &  0.40 &   1.1 &  0.40 & & &\\
304-028 & --1.60&  12.9 & 0.15 &--1.45 &  11.9 &  0.00 & 0.769 &  1.00 & Y\\
...&  0.45 &   4.7 &  0.00 &  0.43 &   3.9 &  0.00 & & &\\
310-032 & --1.75&  19.5 &--0.05 &--1.60 &  11.4 &  0.10 & 0.938 &  0.10 & Y\\
...&  0.35 &   2.1 &  0.50 &  0.43 &   3.5 &  0.20 & & &\\
313-036 & --0.85&   8.8 & 0.21 &--0.70 &  11.7 &  0.20 & 1.296 & --1.00 & Y\\
...&  0.35 &   1.2 &  0.20 &  0.33 &   0.9 &  0.40 & & &\\
328-054 & --1.80&  16.8 & 0.35 &--1.50 &  13.3 &  0.50 & 0.970 & --0.08 & Y\\
... &  0.35 &   5.2 &  0.00 &  0.30 &   1.1 &  0.10 & & &\\
337-068 & --0.80&   2.6 & 0.40 &--1.00 &   6.6 &  0.20 & 1.371 & --1.00 & N\\
... &  0.17 &   1.9 &  0.10 &  0.17 &   3.2 &  0.20 & & &\\
347-154 & --2.25&   8.7 & 0.25 &--1.93 &   9.9 &  0.00 & 0.734 &  1.00 & Y\\
...&  0.10 &   5.7 &  0.20 &  0.47 &   4.0 &  0.20 & & &\\
350-162 & --1.65&   9.8 & 0.35 &--1.52 &  12.4 &  0.20 & 0.914 &  0.24 & Y\\
...&  0.40 &   2.5 &  0.30 &  0.17 &   4.7 &  0.30 & & &\\
365-284 & --1.45&   9.2 & 0.15 &--1.23 &  10.4 &  0.20 & 0.814 &  0.82 & Y\\
... &  0.40 &   3.1 &  0.20 &  0.50 &   2.2 &  0.10 & & &\\
383-318 & --0.55&  10.3 & 0.14 &--0.40 &  10.6 &  0.30 & 1.709 & --1.00 & Y\\
... &  0.15 &   2.2 &  0.10 &  0.12 &   1.3 &  0.30 & & &\\
393-330 & --1.00&  10.0 &--0.05 &--0.80 &  11.9 &  0.00 & 1.522 & --1.00 & Y\\
... &  0.45 &   1.8 &  0.00 &  0.38 &   1.3 &  0.30 & & &\\
398-341 & --0.50&  12.1 & 0.22 &--0.33 &  14.7 &  0.20 & 1.204 & --1.00 & Y\\
... &  0.30 &   1.5 &  0.30 &  0.28 &   3.7 &  0.00 & & &\\
401-344 & --2.25&   9.2 & 0.25 &--2.10 &  15.0 &  0.30 & 0.679 &  1.00 & Y\\
... &  0.40 &   5.8 &  0.30 &  0.15 &   5.1 &  0.20 & & &\\
NB16    & --0.90&   2.0 & 0.53 &--0.95 &   4.6 &  0.40 & 1.232 & --1.00 & N\\
... &  0.15 &   1.4 &  0.10 &  0.22 &   0.8 &  0.10 & & &\\
NB67    & --0.95&   2.2 & 0.54 &--0.90 &   4.7 &  0.50 & 1.273 & --1.00 & N\\
... &  0.15 &   1.4 &  0.30 &  0.10 &   0.3 &  0.20 & & &\\
NB89    & --0.55&   8.3 & 0.24 &--0.65 &  10.4 &  0.20 & 1.128 & --1.00 & Y\\
... &  0.10 &   2.6 &  0.10 &  0.12 &   2.7 &  0.10 & & &\\
\enddata
\tablenotetext{a}{Bruzual \& Charlot (2003) models.}
\tablenotetext{b}{Thomas, Maraston \& Bender (2003) models.}
\tablenotetext{c}{Lee et al. 1994.}

\end{deluxetable}

\end{document}